\documentclass[%
 amsmath,amssymb,
preprint,%
longbibliography
]{revtex4-1}

\usepackage{graphicx}
\usepackage{dcolumn}
\usepackage{bm}
\usepackage{natbib}
\usepackage[colorlinks=true,citecolor=blue,linkcolor=blue,pdfhighlight =/O]{hyperref}
\hypersetup{urlcolor=blue}

\begin{document}


\title[A Scanning Hall Probe Microscope for high resolution, large area, variable height Magnetic Field Imaging]{A Scanning Hall Probe Microscope for high resolution, large area, variable height Magnetic Field Imaging}

\author{Gorky Shaw}
 \email{gorky.shaw@ulg.ac.be.}
 
 \altaffiliation {Present address: University of Liege (ULg), Department of Physics, Experimental Physics of Nanostructured Materials, Sart Tilman, B-4000, Belgium}
 
\author{R. B. G Kramer}
 
\author{N. M. Dempsey}

\author{K. Hasselbach}

\affiliation{ 
	Universit\'{e} Grenoble Alpes, Institut N\'{e}el, F-38042 Grenoble, France
}%
\affiliation{%
CNRS, Institut N\'{e}el, F-38042 Grenoble, France
}%

\date{\today}

\begin{abstract}
We present a Scanning Hall Probe Microscope operating in ambient conditions. One of the unique features of this microscope is the use of the same stepper motors for both sample positioning as well as scanning, which makes it possible to have a large scan range (few mm) in $x$ and $y$ directions, with a scan resolution of 0.1 $\mu$m. Protocols have been implemented to enable scanning at different heights from the sample surface. The $z$ range is 35 mm. Microstructured Hall probes of size 1-5 $\mu$m have been developed. A minimum probe-sample distance $<$ 2 $\mu$m has been obtained by the combination of new Hall probes and probe-sample distance regulation using a tuning fork based force detection technique. The system is also capable of recording local $B(z)$ profiles. We discuss the application of the microscope for the study of micro-magnet arrays being developed for applications in micro-systems.
\end{abstract}

\maketitle


\section{Introduction}
\label{secIntro}
Scanning Hall Probe Microscopy (SHPM) is a popular local magnetic characterization technique \cite{Chang1992,Oral1996,Oral1996a,Howells1999,Sandhu2002}. Among common non-invasive magnetic imaging techniques, SHPM is particularly useful due to a number of advantages. In comparison to Magnetic Force Microscopy (MFM) and Magneto-Optical Imaging (MOI), SHPM enables direct quantitative mapping of magnetic field. MFM is more challenging for quantitative measurements as characterization and modeling of the magnetization of the tip are rather difficult \cite{Bending2010}. In comparison to MOI, SHPM works in a much larger magnetic field range, and offers better spatial resolution in the case when the sample is not completely flat \cite{Kirtley2010}. Though SHPM involves longer image acquisition time than MOI unlike in a scanning probe system, in MOI the whole image is acquired at one time, and involves acquisition times of a few seconds at most) and offers worse spatial resolution compared to MFM, the unique compromise between spatial resolution and field sensitivity makes it a very useful tool for convenient non-destructive quantitative characterization over wide temperature and field ranges. Over the years, efforts have been made to improve the spatial resolution and scan range \cite{Gregory2002,Dinner2005,Cambel2005,Tang2014}, temperature range \cite{Karci2014}, and magnetic field resolution \cite{Shimizu2004} of SHPM.

In this paper we report on a room temperature SHPM system developed to characterize arrays of micro-magnets. SHPM is the tool of choice for quantifying the stray fields produced by micro-magnets, which have many applications in bio-medical studies \cite{Zanini2011,Brunet2013,Pivetal2014} and MEMS \cite{Dempsey2009}. The present SHPM system demonstrates the usefulness of a combination of concepts, tuning fork feedback based height control, monolithic electronics for regulation and Hall signal read out, scanning with stepper motors, compared to microscopes presented before\cite{Gregory2002,Dinner2005,Cambel2005,Tang2014,Karci2014,Shimizu2004,Kustov2010}. Microstructured Hall probes with active area of size 1-5 $\mu$m have been fabricated for use in the microscope. The characteristics of the microscope include: large in-plane scan range (up to a few mm, limited by sample topography) with fine scan resolution (step resolution of the motors is 0.1 $\mu$m, which is, in principle, the minimum possible scan resolution. However, the spatial resolution of the magnetic field distribution obtained in this technique is limited by the size of the Hall probes as the detected magnetic field is the average field over the active area of the probe) and large $z$ range, allowing sample-probe distances from  $<$ 2 $\mu$m to 35 mm, with a step resolution of 0.2 $\mu$m, as well as large magnetic field detection (fields up to $\sim$ 1 T have been measured) with high field resolution (100 $\mu$T). Apart from scanning, the system is also capable of fast recording of local $B(z)$ profiles. In the following sections, we first describe the microscope. This is followed by some results showcasing the capabilities of the microscope.


\section{The SHPM system}
\label{secSystem}
\subsection{The Microscope}
\label{subsecMicroscope}
\begin{figure*}
	\centering
	\includegraphics[width=16.0cm]{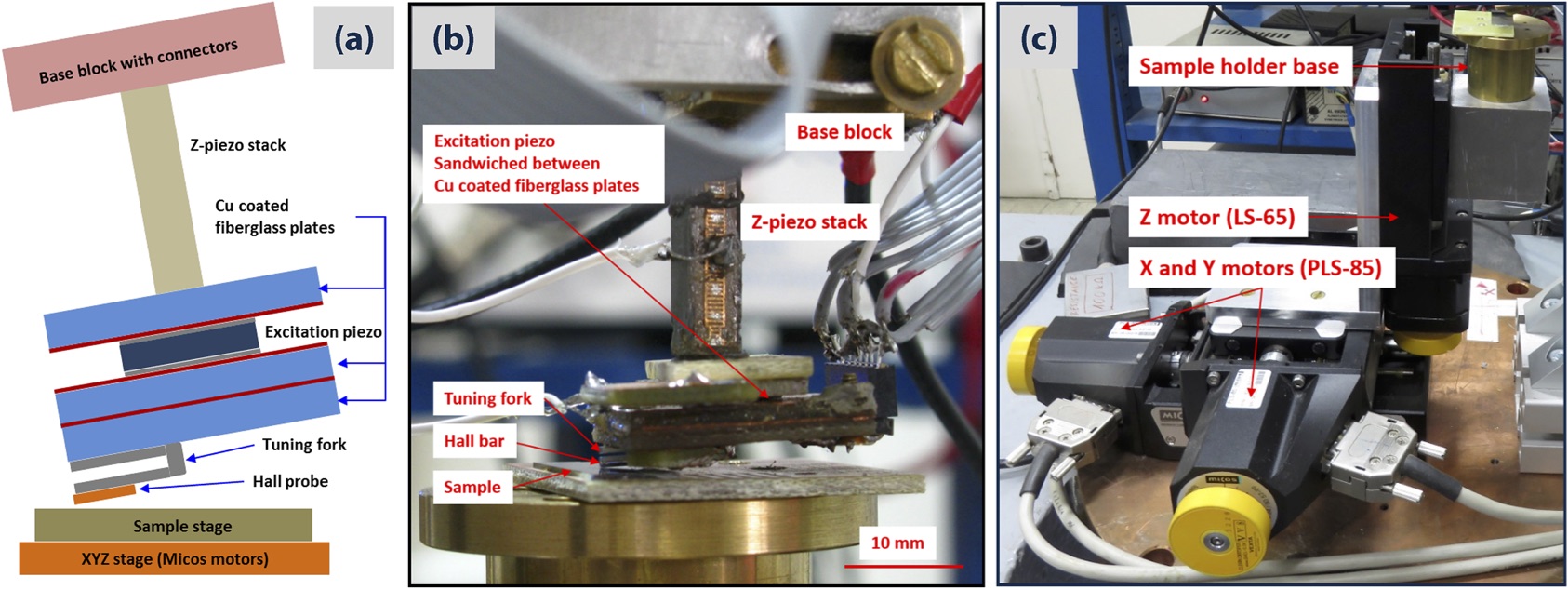}
	\caption{(a) Schematic of the microscope. (b) Optical image of the microscope showing the probe and sample stages. (c) The sample stage showing the Micos motors.}
	\label{f1}
\end{figure*}

\begin{figure*}
	\centering
	\includegraphics[width=13.0cm]{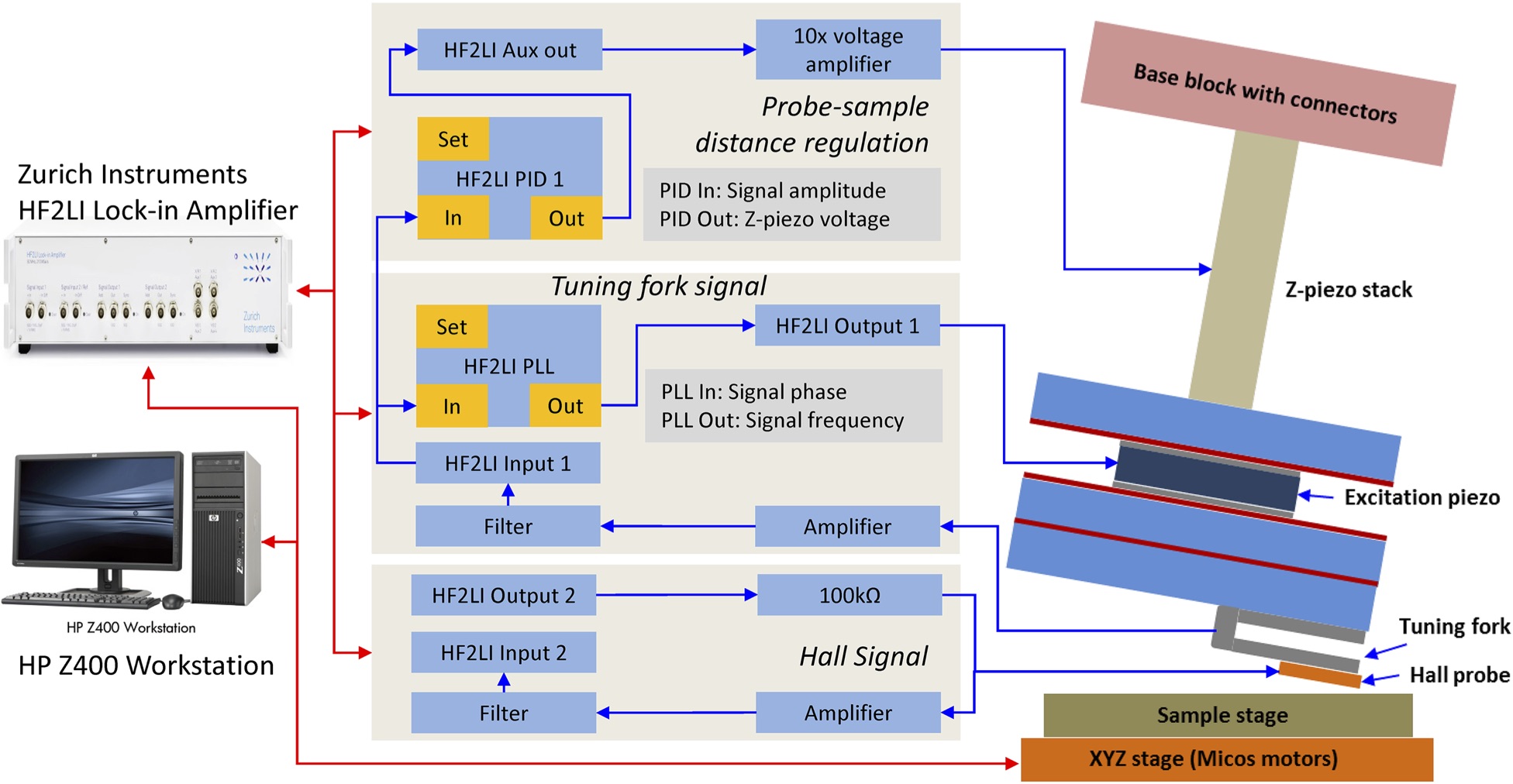}
	\caption{Schematic diagram of the SHPM setup.}
	\label{f2}
\end{figure*}

Figures \ref{f1}(a) and \ref{f1}(b) show a schematic and an optical image, respectively, of the microscope. The microscope is mounted on a vibration-isolation table. Vibrations from the floor are filtered by a concrete block supported by rubber dampers filled with compressed air flowing viscously between the dampers and air reservoirs. The microscope is mounted on a copper block that rests on the concrete block. The microscope consists of two separate parts, viz., a mobile part which consists of a sample holder placed on an XYZ stage, and a fixed part which consists of the probe mounted on a tuning fork. This second part includes piezoelectric elements for $z$-regulation, tuning fork excitation, and all electrical connections involved in the measurement. Note that it is important that both of these parts rest on the same base (in this case the copper block) as otherwise additional and uncontrolled tilt between the sample and probe planes would arise which would lead to unnecessary complications in sample-probe distance regulation. The microscope is covered with a 50 $\times$ 50 $\times$ 60 cm$^{3}$ opaque enclosure to minimize the influence of ambient light. Figure \ref{f2} shows a schematic of the complete setup. A fast lockin-amplifier (Zurich Instruments HF2LI, henceforth referred to as \textquoteleft the HF2LI\textquoteright) \cite{zhinst} serves as the primary electronics component of the microscope. It is used to provide the excitation signal as well as to measure the current from the tuning fork. It is also used to provide the Hall probe excitation current as well as to measure the generated Hall voltage. Finally, it also provides the PID control for the $z$-piezo stack mentioned (explained below). All control and measurement operations are performed through custom-built reliable and user-friendly LabVIEW programs developed specifically for the setup.

\subsubsection{Sample stage}
\label{subsubsecSampleStage}
The mobile sample stage consists of the sample holder base mounted on the primary scanning component which is a Micos stepper motor stage as shown in Figure \ref{f1}(c). The sample can be moved in all three directions (X, Y, Z) by virtue of the stepper motors. The three Micos motors used are: two linear Stages \textquotedblleft Precision Linear Stage PLS-85\textquotedblright \cite{pls85} for X and Y motion, and one linear stage \textquotedblleft Linear Stage LS-65\textquotedblright \cite{ls65} for Z motion.The travel range in X and Y directions is 50 mm and the minimum step size is 0.1 $\mu$m. This minimum step size is precise enough for the primary purposes the microscope is built for. In our microscope, the stepper motors themselves perform the scanning, additional piezo scanners are superfluous. As a result, in principle, the maximum achievable scan range is the same as the travel range of the X/Y motors = 50 mm. However, in practice, the effective scan range is limited by the properties of the sample being measured. Most often, irregularities of the sample surface mean that the probe-sample distance regulation (discussed in Section \ref{subsubsecDistRegulation}) can be maintained only over a smaller distance, thus limiting the scan range. We have been able to perform scans over areas as large as 2.5 mm $\times$ 2.5 mm (cf. results of our measurements on bulk magnets in Section \ref{subsecBulkMagnets} showing 2 mm $\times$ 2 mm images). For comparison, minimum step size and maximum possible scan range at room temperature of some other reported stepper motor based long-range SHPM systems are as follows: 1.25 $\mu$m and 25 mm $\times$ 25 mm in Ref. \citenum{Gregory2002}, 0.2 $\mu$m and 1 cm $\times$ 4 cm in Ref. \citenum{Dinner2005}, 5 $\mu$m and 7 mm $\times$ 25 mm in Ref. \citenum{Cambel2005}, and 0.1 $\mu$m and 1 mm $\times$ 1 mm in Ref.\citenum{Tang2014}. All three motors in our setup are characterized by very reliable repeatability (negligible backlash error). The linear stages are controlled by a \textquotedblleft SMC Corvus PCI\textquotedblright controller \cite{corvus} installed in the PC. The system has a fast response, allowing stepping at 50 ms intervals. At this stepping interval, a typical scan (200 $\times$ 200 pixel) takes about two hours with additional overheads (cf. Section \ref{subsubsecScanProtocol} for details on the scan protocol). In comparison, in the SHPM system described in Ref. \citenum{Dinner2005}, a 200 $\times$ 200 pixel scan takes several hours, and in Ref. \citenum{Gregory2002} a 60 $\mu$m $\times$ 60 $\mu$m scan with pixel size 1.25 $\mu$m takes about 90 minutes.

\subsubsection{Probe stage}
\label{subsubsecProbeStage}
The fixed probe stage consists of the Hall probe glued to a commercial quartz tuning fork \cite{quartz} which is in turn glued to a probe plate (Cu coated fibre glass plate). Electrical connections to the Hall probe and tuning fork are anchored on the probe plate. The tuning fork is mechanically excited through a \textquoteleft thickness-mode\textquoteright \space piezoceramic element below the tuning fork (labelled \textquoteleft Excitation piezo\textquoteright \space in Fig. \ref{f1}(a)) \cite{Veauvy2002,Hykel2014}. The excitation signal is provided from the first output of the HF2LI (cf. Figure \ref{f2}). This assembly is attached to a \textquoteleft $z$-piezo stack\textquoteright\space extension piezoelectric element \cite{zpiezo} which is used for regulation of the probe-sample distance. The $z$-piezo stack has a vertical extension range $\sim$ 32 $\mu$m (with input voltage range 0-120 V). Voltage on the $z$-piezo stack is applied through one of the auxiliary outputs of the HF2LI via a low noise voltage amplifier with 10x amplification (cf. Figure \ref{f2}). The piezo stack is mounted on a base block that can be tilted in order to adjust the angle between the Hall probe and the sample surface.

\begin{figure}
	\centering
	\includegraphics[width=6 cm]{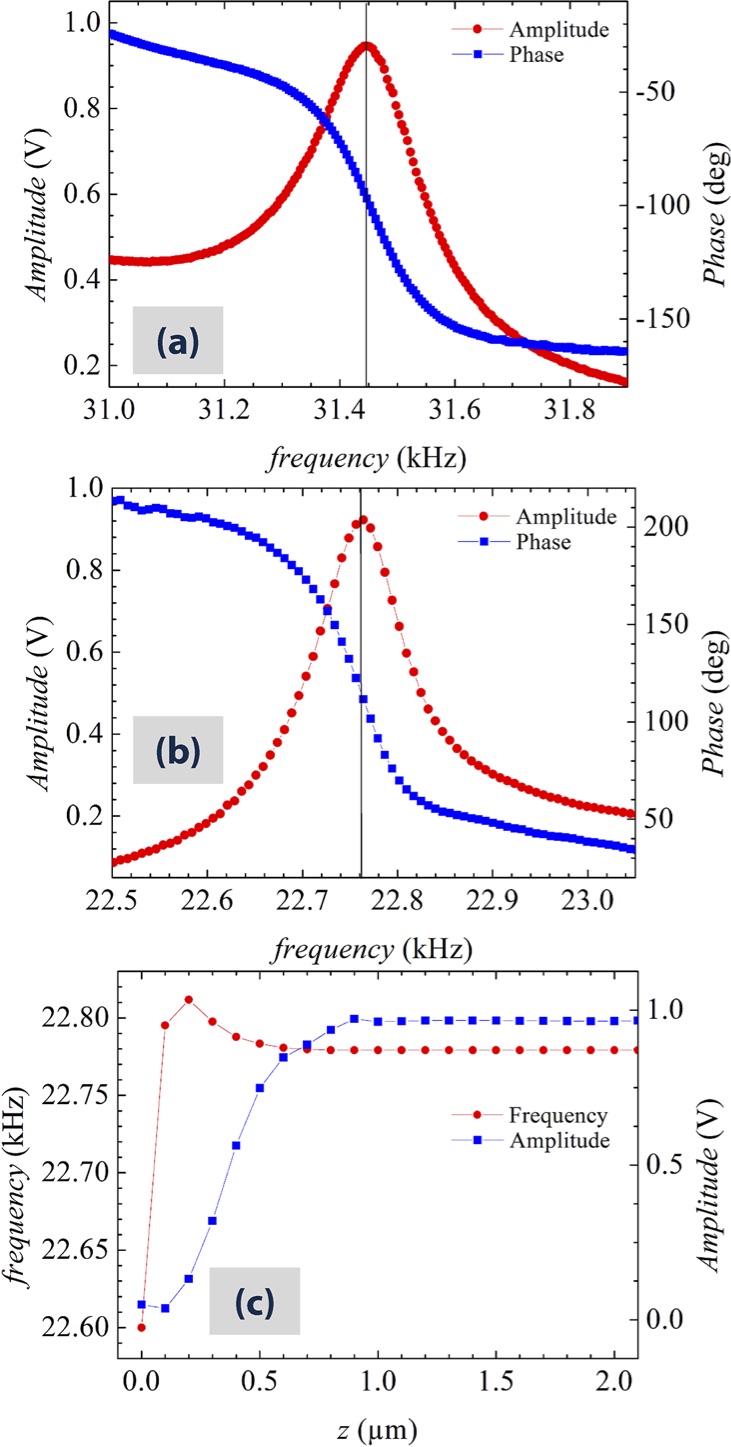}
	\caption{(a) and (b) Frequency response of a tuning fork glued to a probe plate, without and with a probe mounted, respectively. (c)Approach curve of a tuning fork with a probe mounted on it (cf. text for details).}
	\label{f3}
\end{figure}

\subsubsection{Probe-sample distance regulation}
\label{subsubsecDistRegulation}
One of the primary challenges in a scanning probe setup is the ability to approach the probe as close as possible to the sample surface without crashing the probe into the sample. This is achieved by real-time control of the probe-sample distance by regulating the extension of the $z$-piezo stack during a scan using the amplitude of vibration of the tuning fork for feedback. Tuning fork feedback based probe-sample distance control was first proposed by Khaled Karrai and Robert D. Grober \cite{Karrai1995,Karrai1995a} and has since been widely used in near-field scanning optical microscopy (NSOM). This technique has been implemented in Scanning SQUID \cite{Veauvy2002,Hykel2014} and Scanning Hall probe \cite{Dede2008} Microscopes. The voltage on the $z$-piezo stack is regulated by a fast closed-loop Proportional-Integral-Derivative (PID) control in the HF2LI PID settings have been optimized based primarily on the sample surface type. Typical values for samples with hard surfaces (e.g., thermomagnetically patterned micromagnets) are: P = $-0.05$, I = $-50$ s$^{-1}$, D = $+1$ $\mu$s. For samples with soft surfaces (e.g., topographically patterned micromagnets embedded in polymer) typical values of the parameters are: P = $-0.5$, I = $-500$ s$^{-1}$, D = $+4$ $\mu$s. These settings correspond to an integral time constant (P/I) = 1 ms. The PID output is updated at 100 kHz. The input of the PID is the amplitude of vibration of the tuning fork (cf. Figure \ref{f2}). Current proportional to the amplitude of vibration is generated in the tuning fork, which is amplified with a current/voltage converter of gain 10$^{7}$ V/A, and passed through a high-pass filter (cut-off frequency 8 kHz). The amplified voltage is demodulated by the HF2LI and the phase signal is fed into a phase locked loop (PLL). The PLL ensures that the tuning fork oscillation frequency always remains in the vicinity of the resonance frequency. The excitation frequency is adjusted in order to maintain the tuning fork at resonance (cf. Figure \ref{f2}). The resonance frequency $f_{R}$ of the free tuning fork is $\sim$ 32.768 kHz. However, it depends strongly on environmental conditions. Figure \ref{f3}(a) shows a typical frequency response of a tuning fork glued to the probe plate. Note that $f_{R}$ is reduced to $\sim$ 31.45 kHz. Figure \ref{f3}(b) shows the frequency response after a probe is glued to the tuning fork. In this case $f_{R}$ is further reduced to $\sim$ 22.76 kHz. $f_{R}$ for a tuning fork with a probe attached varies in the range 14-23 kHz. Note that the amplitude of oscillation of the tuning fork also reduces with increase in load. This is not obvious from Figures \ref{f3}(a) and \ref{f3}(b) as the excitation voltage in each case is adjusted such that the amplitude at resonance is always $\sim$ 1 V. 

Further, the amplitude of oscillation monotonically decreases on approaching the sample surface. This is the basis of the probe-sample distance regulation \cite{Karrai1995,Karrai1995a}. Figure \ref{f3}(c) shows a typical approach curve for a loaded tuning fork (i.e., after a probe is mounted on the tuning fork glued to a probe plate). The $x$-axis shows the distance of the tuning fork from the sample surface (sample surface at $z$ = 0). Note from Figure \ref{f3}(c) that as the probe approaches the sample surface, the resonance frequency increases before decreasing, while the amplitude of oscillation decreases monotonically. Hence it is possible to maintain a fixed desired probe-sample distance if the amplitude of oscillation can be regulated. This is implemented via the PID control of the extension of the $z$-piezo stack using the amplitude of oscillation of the tuning fork as input as discussed above. In general, 70\% of amplitude of oscillation of the tuning fork when it is far away from the sample surface is used as the desired set-point for probe-sample contact. For samples with rough and/or soft surfaces, 75-80\% is set. From Figure \ref{f3}(c), z $\approx$ 470 nm for 70\% and 500 nm for 80\%. This is the probe-sample distance which is maintained during a scan in the regulation mode (cf. Section \ref{subsubsecScanProtocol}).

\subsection{Magnetic field detection}
\label{subsecFieldDetect}
\begin{figure*}
	\centering
	\includegraphics[width=10 cm]{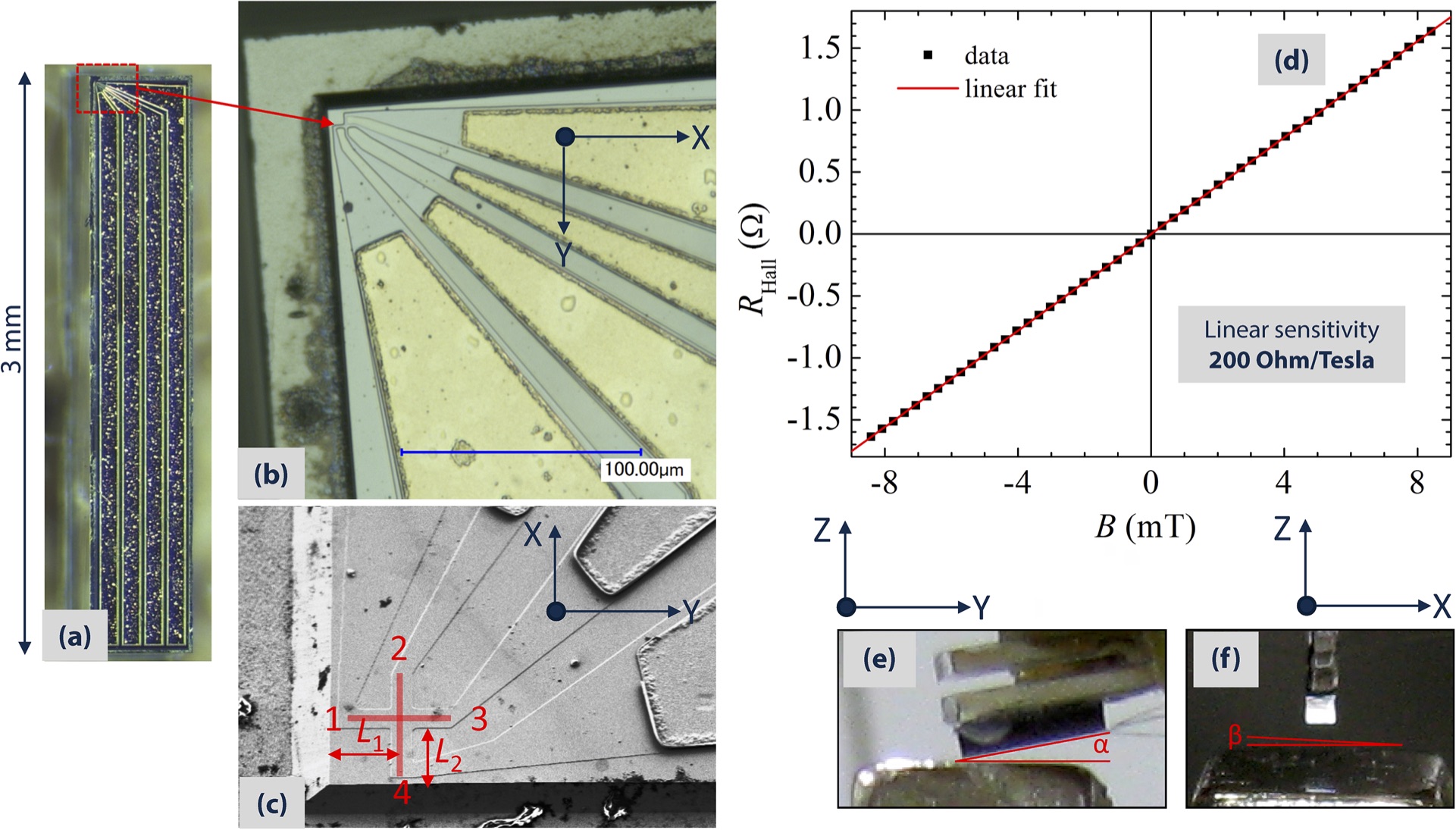}
	\caption{(a) Optical image of a Hall bar. (b) Optical image of a Hall bar showing the region of the Hall bar marked by the red rectangle in (a). The Hall cross is indicated by the red arrow. (c) SEM image of the Hall bar showing the Hall cross and leads. The Hall cross is indicated by the red lines. The distances of the Hall cross from the edges of the Hall bar are labelled as $L_{1}$ and $L_{2}$. (d) Typical calibration curve of a Hall probe showing $B$ vs. $R_{Hall}$ data and linear fit of the data. (e) and (f) Side and front views, respectively, of the probe, showing the angles $\alpha$ and $\beta$ the probe makes with the sample surface.}
	\label{f4}
\end{figure*}

\subsubsection{Hall probes}
\label{subsubsecHallProbes}
The Hall probes consist of heterostructures of GaAs/AlGaAs two-dimensional electron gas (2DEG) materials. The two dimensional electron gas is about 50 nm below the surface. The mesa is first etched with a chemical etch by 100 nm in order to structure the Hall cross. Then a multilayer of Ni/Au/Ge is evaporated with an e-gun evaporator (Plassys). Rapid thermal annealing allows the diffusion of the AuGe to the two-dimensional electron gas, thus forming the Ohmic contacts with the 2DEG. An additional deep etch $\sim$ 10 $\mu$m is made in order to place the Hall cross exactly at the corner of the wafer. Finally, the Hall probes are cut with a diamond dicing machine. The electron sheet density $n$ is $\sim 3 \times 10^{12}$ cm$^{-2}$ (calculated from the relation $n$ = 1/($ke$), where $e$ is the electronic charge and $k$ is the Hall sensitivity = 200 $\Omega$/T, cf. Section \ref{subsubsecProbeCalibMagMeas} below). The mobility of the electron gas measured at 4.2 K is $\sim$ 30000 cm$^{2}$/(V·s). Figure \ref{f4}(a) shows an optical image of a typical Hall bar. The optical image in Figure \ref{f4}(b) shows a close-up of the Hall bar with the Hall cross indicated by the red arrow. Figure \ref{f4}(c) shows an SEM image of the Hall bar showing the Hall cross and contact leads. The Hall cross is identified by the red lines drawn on the image. The four edges of the Hall cross are marked as 1, 2, 3 and 4. The distances of the Hall cross from the edges of the Hall bar are shown as $L_{1}$ and $L_{2}$. Hall probes of size 1-5 $\mu$m have been prepared. The particular probe shown here is of size $\sim$ 4 $\mu$m. Resistance values through the probes were found to be $\sim$ 15-17 k$\Omega$ for the smaller probes and 5-6 k$\Omega$ for the larger probes. Though Hall probes of smaller size have been prepared with GaAs/AlGaAs, smaller sizes are also associated with lower field sensitivity\cite{Chang1992,Sandhu2002,Hicks2007}, hence the size of the present probes is suitable for our purposes.

\subsubsection{Probe calibration and magnetic field measurement}
\label{subsubsecProbeCalibMagMeas}
The Hall probe is excited with an ac signal from the second output of the HF2LI through a 100k$\Omega$ resistor in series. The Hall excitation current is set in the range 10-50 $\mu$A (pk-pk). The frequency of excitation is set at 522 Hz. This was chosen empirically to correspond to minimum noise in the output signal. The amplified and filtered Hall voltage is detected at the second input of the HF2LI (cf. Figure \ref{f2}). A number of Hall probes have been used in the measurements. All the Hall probes show remarkably consistent magnetic field sensitivity. The Hall probes were calibrated using a pre-calibrated copper coil magnet (16.83 G/A, or 1.683 mT/A) \cite{KustovThesis}. For calibration, a Hall probe is placed at the center of the magnet and the Hall voltage is measured for a set of values of current flowing through the solenoid (corresponding to magnetic field values in the range -8 mT to +8 mT, in the magnet). Figure \ref{f4}(d) shows typical calibration data for one of the Hall probes. The $y$-axis shows the Hall resistance $R_{Hall}$.
\begin{equation}
\label{eqn1}
R_{Hall}=V_{13}/I_{24}
\end{equation}
Where $I_{24}$ is the current applied across the sides 1 and 3 of the Hall cross and $V_{13}$ is the Hall voltage (in-phase component of the demodulated signal) measured across the sides 2 and 4. From Figure \ref{f4}(d) it is clear that the Hall resistance varies very linearly with magnetic field. From the linear fit, the Hall sensitivity ($k$) = is found to be 200 $\pm$ 2 $\Omega$/T for all the used probes, which is consistent with the fact that Hall probe sensitivity is nearly independent of the Hall cross area \cite{Kirtley2010}. In our Hall probes the doping level is rather high which results in the relatively lower Hall sensitivity compared to some other reported values at room temperature (300 K) for GaAs/AlGaAs heterostructures (3000 $\Omega$/T in Ref. \citenum{Oral1996}, 2000 $\Omega$/T in Ref. \citenum{Tang2014}, 1750 $\Omega$/T in Ref. \citenum{Dede2008}) however, it is sufficient for our purposes. We believe this also ensures that conduction always takes place preferentially in the 2DEG as there are many charge carriers available. Indeed, the Hall sensitivity was measured at room temperature as well as down to 50 mK and is found to be nearly constant in this temperature range (variation smaller than 5\%). 

Once the sensitivity of the Hall probe is known, the measured Hall voltage $V_{13}$ can be simply converted to magnetic field $B_{z}$ as,
\begin{equation}
\label{eqn2}
B_{z} = V_{13}/I_{24}/k
\end{equation}
Or,
\begin{equation}
\label{eqn3}
B_{z} = R_{Hall}/k
\end{equation}
 Note that there is always some offset voltage, i.e., a non-zero measured Hall voltage even in zero field. This can appear due to various reasons, including misalignment of the voltage leads and/or leakage between the leads. The offset voltage is measured before each measurement and the magnetic field value is determined by subtracting this offset signal from the measured Hall signal.

The magnetic field resolution is limited by the noise in the output signal. The noise depends on the active area of the probe and is larger for a smaller probe. In order to obtain understanding of the sensitivity limitations of our microscope let us compare the observed noise level to the Johnson-Nyquist thermal noise which limits the sensitivity of measurements \cite{Oral1996,Geim1997,Novoselov2003}:
\begin{equation}
\label{eqn4}
V_{johnson} = \sqrt{4k_{B}TR}
\end{equation}

Where $k_{B}$ is the Boltzmann constant, $T$ is the temperature and $R$ is the resistance of the Hall sensor. At room temperature ($T$ = 300 K), using $R$ = 120 k$\Omega$ (maximum contact resistance for the smallest probe, including the series resistance 100 k$\Omega$), $V_{johnson} \sim$ 1.424 nV/$\surd$Hz. For excitation current of 10 $\mu$A, this gives a lowest possible noise in $B_{z}$ of $\Delta B_{z} \sim$ 1 $\mu$T/$\surd$Hz, or a magnetic field sensitivity of 10 $\mu$T (at measurement frequency 522 Hz and time constant 10 ms). However, the magnetic field resolution in our measurements is $\sim$ 100 $\mu$T. Thus additional noise sources other than the Johnson noise are at the origin of this field resolution. A noise floor of $\sim$ 16 nV/$\surd$Hz is measured when the Hall probe is not excited. It increases to $\sim$ 18 nV/$\surd$Hz when the probe is excited. The measured noise level is thus essentially due to the wiring of the probe, capturing ambient electrical noise, as the microscope operates in an unshielded environment. The field sensitivity in our system is in between values reported for SHPM systems operating at room temperature (300 K): 0.08 $\mu$T/$\surd$Hz in Ref. \citenum{Gregory2002} (InSb Hall probes), 350 $\mu$T/$\surd$Hz in Ref. \citenum{Tang2014} (GaAs/AlGaAs Hall probes). 

\subsection{Measurement protocol}
\label{subsecMeasProtocol}
\subsubsection{Estimate of distance of probe from sample surface}
\label{subsubsecHeightEstimate}
For proper interpretation of experimental data, it is important to reliably estimate the distance between the Hall cross and the sample surface. To avoid contact of the wires micro-bonded to the Hall probe with the sample surface and for safety of the probe against damage by hitting the sample surface, the probe is kept at a slight inclination w.r.t. the sample stage. Figures \ref{f4}(e) and \ref{f4}(f) show side and front views, respectively, of a probe mounted on a tuning fork with a sample below, indicating the angles $\alpha$ and $\beta$ the probe makes with the sample surface. Note that the Hall cross is not exactly at the edge of the bar but at a distance from it. In Figure \ref{f4}(c) the distances of the Hall cross from the Hall bar edges are indicated as $L_{1}$ and $L_{2}$. For the smallest probes (1 $\mu$m size), $L_{1}$ = $L_{2}$ = 8 $\mu$m. For the largest probes (5 $\mu$m size), $L_{1}$ = $L_{2}$ = 15 $\mu$m. The angle $\alpha$ ensures that the probe does not hit the sample surface and the angle $\beta$ ensures that the edge of the Hall bar closer to the Hall cross touches the sample surface. The angle $\alpha$ is generally set at $\sim$ 9-10$^{\circ}$ ($\pm$ 0.5$^{\circ}$, measured optically), which is small enough to assume that the measured signal the component of the magnetic field is perpendicular to the sample surface \cite{KustovThesis}. Note that $\alpha \sim$ 5$^{\circ}$ is usually good enough to ensure that the micro-bonded wires do not touch the sample surface. However, a larger angle is needed for safety of the probe itself. Since the Hall cross is rather close to the edge of the Hall bar, scanning over large areas of samples with large topographic variations leads to significant bruising of the probe. While scanning with a smaller angle $\alpha$, in a number of probes, the contact closest to the edge got disconnected due to scratches induced during scanning, rendering the probe unusable. The angle $\beta$ is kept as small as possible ($\sim$ 1$^{\circ}$). The distance $h$ of the Hall cross from the sample surface is then given by,
\begin{equation}
\label{eqn5}
h = L_{1} sin\alpha + L_{2}sin\beta cos\alpha
\end{equation}
Using $L_{1}$ = $L_{2}$ = 8 $\mu$m (15 $\mu$m), $\alpha$ = 9$^{\circ}$ and $\beta$ = 1$^{\circ}$ in Equation \ref{eqn5}, we get, $h \approx$ 1.4 $\mu$m (2.6 $\mu$m) for the smallest (largest) probe, which is the minimum possible probe-sample distance that can be achieved without endangering the probe. 

The distances of the Hall cross from the Hall bar edge are comparable to some other reported SHPM systems (13 $\mu$m\cite{Oral1996}, 20 $\mu$m\cite{Tang2014}) and are significantly smaller than that in the probes used in the previous setup\cite{Kustov2010} (100 $\mu$m). As evident from the discussion above, the smaller distance in the present Hall probes has allowed achieving a much smaller probe-sample distance in the present setup $\sim$ 2 $\mu$m as compared to $\sim$ 25 $\mu$m in the previous setup). 

\subsubsection{Scan protocol: regulation and flyover modes}
\label{subsubsecScanProtocol}
The scanning plane is designated as the $(x, y)$ plane and hence the measured magnetic field component is along the $z$-axis ($B_{z}$). As discussed in Section \ref{subsubsecDistRegulation}, the probe is mounted on a quartz tuning fork allowing topographic feedback to maintain constant probe - sample distance during a scan. After setting the desired starting $(x, y)$ position, the sample is approached towards the probe with active topographic feedback via the PID control, till the desired probe-sample distance is reached ($<$ 500 nm, determined by the set amplitude of oscillation of the tuning fork, usually 70\% of the amplitude of oscillation when the tuning fork is far away from the sample surface, cf. Section \ref{subsubsecDistRegulation}). The fast scan direction is along the $y$-axis, i.e., the long side of the Hall bar (cf. Figure \ref{f4}). Scans are performed in the $y$-direction after setting the desired $x$ before each scan in $y$. While making a step in $y$ during the scan, the probe - sample distance is continuously regulated through the fast PID control. At each $y$-step, the magnetic field signal and sample topography information are recorded simultaneously. Measurements are always made while scanning in the same direction, i.e., positive $y$ $(y+)$ direction in order to avoid any appearance of effects of the backlash of the motors. While returning in $y$ $(y-)$, PID control is turned off and the $z$-piezo is fully retracted for safety. The next $x$-step is also made with the $z$-piezo fully retracted and PID control off. Then the PID control is turned on again and the next scan along $y+$ is performed. In the resulting SHPM images, the $(x+)$ direction is from left to right and $(y+)$ direction is from top to bottom. Note that it is the sample which is moved during the scan while the probe remains stationary.

In our setup, apart from this \textquoteleft regulation mode\textquoteright, scanning at a fixed height from the sample surface in \textquoteleft flyover mode\textquoteright, without such feedback, is also possible. Typically, in a particular measurement, a first scan is performed in the regulation mode, with the probe close to the sample surface. Next, the probe is retracted to the desired height from the sample surface and a scan is performed without topographic feedback for regulation of the extension of the $z$-piezo stack. Instead, the extension is determined from the topography information recorded earlier in the regulation mode. In this way, a set of scans can be performed at various heights from the sample surface. Note that this flyover mode of scanning allows to follow the topography and tilt of the sample. This is unlike standard lift-off mode scanning where the probe is lifted to a fixed height above the sample surface and scanning is performed in a flat plane without taking into account the sample topography and tilt. Standard lift-off mode scanning in a flat horizontal or tilted plane is also possible in our microscope. It is possible to perform scans at a height of up to 35 mm from the sample surface (range of the Z-motor). 

\subsubsection{$B(z)$ profiles}
\label{subsubsecB(z)Profiles}
Apart from performing scanning measurements, the system is also capable of, and has been extensively used for, fast recording of local $B(z)$ profiles. For these measurements, the probe is approached close to the sample surface at the desired $(x, y)$ location. Then the sample is retracted to a set probe-sample distance while the Hall signal is recorded at specified $z$ intervals. In this way, local $B(z)$ profiles at different desired $(x, y)$ locations can be obtained.


\section{Measurements}
\label{secMeasurements}
The SHPM setup has been extensively used for quantitative characterization of a variety of magnetic structures, both hard and soft magnetic materials, topographically as well as thermomagnetically patterned micro-magnetic structures, and bulk permanent magnets. Discussed in the following sections are some of the experimental results highlighting the various capabilities of the microscope.

\begin{figure}
	\centering
	\includegraphics[width=8cm]{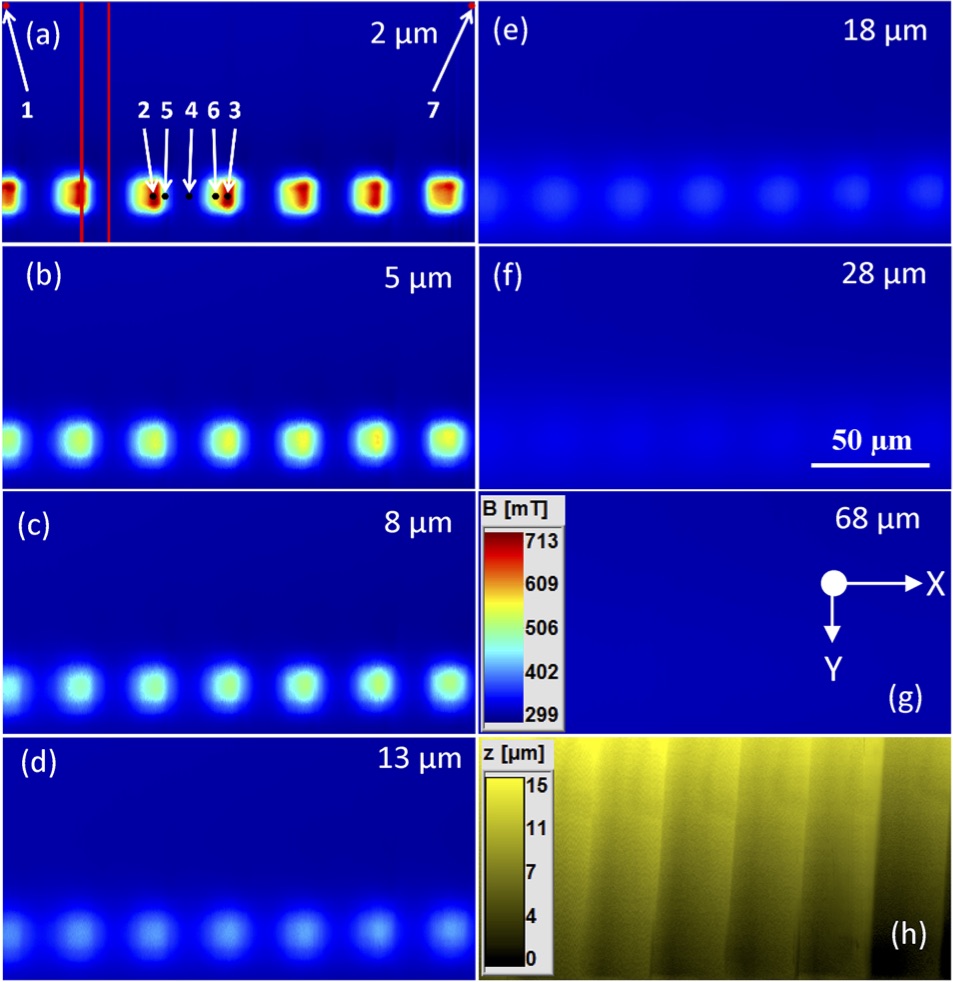}
	\caption{(a)-(g) SHPM showing the $B_{z}$ distribution in a 1D array of  topographically patterned Fe-Co soft micro-magnet pillar structures embedded in a polymer (cf. text for details). The image in (a) is obtained in the regulation mode and those in (b)-(g) are obtained in the flyover mode (cf. text for details). The height of the probe from the sample surface corresponding to each image is indicated in the respective panels. The inset in (g) shows the common scale bar. (h) Image showing topography of the sample. It is obtained simultaneously along with the SHPM image during the regulation mode scan.}
\label{f5}
\end{figure}
\begin{figure}
	\centering
	\includegraphics[width=6 cm]{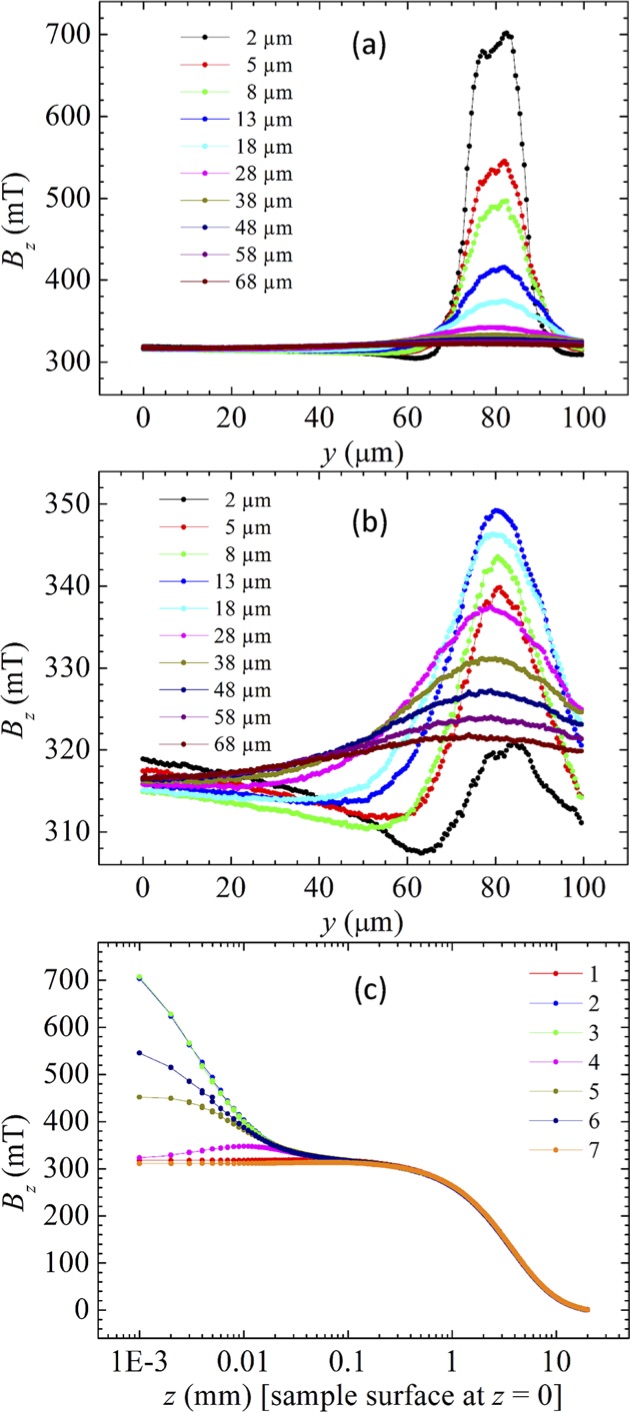}
	\caption{(a) and (b) $B_{z}(y)$ profiles at different scan heights along the red lines shown in Figure \ref{f5}(a), one profile along a region with high field gradients and one along a region with low field gradients. (c) $B_{z}(z)$ profiles at different locations on the image as indicated by numbers in Figure \ref{f5}(a).}
	\label{f6}
\end{figure}

\subsection{1D array of topographically patterned soft micro-magnets}
\label{subsecFeCoPillars}
Figures \ref{f5}(a)-(g) show a set of SHPM images showing the $z$-component of the stray field pattern $(B_{z})$ in a 1D array of topographically patterned Fe-Co soft micro-magnet pillar structures positioned above a bulk permanent magnet. Details of the preparation and characterization of the samples can be found in Ref. \cite{Roy2016}. The scan resolution in Figures \ref{f5}(a)-(g) is 0.5 $\mu$m and the scan area = 400 $\times$ 200 pixel = 200 $\mu$m $\times$ 100 $\mu$m. The $B_{z}$ color scale in the images is indicated by the common scale bar in the inset of Figure \ref{f5}(g). Each micro-pillar is 10 $\mu$m wide and the separation is 30 $\mu$m (center-to-center). The pillars are 40 $\mu$m tall. For these measurements, the sample is placed on top of a bulk magnet which produces a field of $\sim$ 320 mT normal to the sample plane at the sample surface. The first image in Figure \ref{f5}(a) was obtained in the regulation mode with the probe-sample distance estimated at $h \sim$ 2 $\mu$m ($\pm$ 0.5 $\mu$m). Then successive images were obtained at different probe-sample distances in the flyover mode up to $h \sim$ 68 $\mu$m. The $h$ corresponding to each image is indicated in the respective panels. The topography recorded in the regulation mode is shown in Figure \ref{f5}(h). The Fe-Co pillars are coated with a layer of PDMS that smoothens the large topographical variations in the sample due to the micro-pillars and hence enables safe scanning of the Hall probe over the sample surface. In Figure \ref{f5}(h) modulations in the PDMS layer due to the underlying micro-pillars, along with an overall tilt in the sample plane w.r.t. the scanning plane, can be observed. From Figures \ref{f5}(a)-(g), strong field modulation, with large flux concentration at the micro-pillars can be clearly observed, with $B_{z}$ as high as 713 mT measured 2 $\mu$m above the sample surface at the position of the pillars (cf. Figure \ref{f5}(a) and scale bar in inset of Figure \ref{f5}(g)). In Figure \ref{f5}(a), a corresponding decrease in flux density in the vicinity of the pillars can also be observed. With increasing probe-sample distance $h$, the field modulation due to the micro-pillars can be seen to reduce. Finally, at $h \sim$ 68 $\mu$m, modulations completely disappear, and a uniform field distribution across the image, corresponding to the underlying bulk magnet, is observed. 

These observations are further elaborated in Figure \ref{f6}. Figures \ref{f6}(a) and \ref{f6}(b) show $B_{z}(y)$ profiles obtained from the set of images shown in Figure \ref{f5} along the two red lines drawn over the image in Figure \ref{f5}(a), through the center of a pillar and in between two pillars, respectively.The height $h$ corresponding to each $B_{z}(y)$ profile is indicated in both panels. The strong flux concentration at the micro-pillars is very clear from Figure \ref{f6}(a). Figure \ref{f6}(b) clearly shows the dip in $B_{z}$ close to the pillars. Further, $B_{z}(z)$ profiles were recorded at various locations of the image area as indicated by numbers in Figure \ref{f5}(a). The $B_{z}(z)$ profiles are shown in Figure \ref{f6}(c). From Figure \ref{f6}(c), the strong flux concentration at the micro-pillars ($>$ 700 mT near the sample surface at positions 2 and 3, on top of micro-pillars) over and above the field due to the bulk magnet ($\sim$ 320 mT near the sample surface at positions 1 and 7, far from the micro-pillars) is very clear. Further, the $B_{z}(z)$ profiles show that the field modulation effect of the pillars completely disappears beyond $z \sim$ 0.1 mm, where all the profiles merge into the profile expected for the bulk magnet only.

\subsection{Hard magnetic powder based micro-flux sources}
\label{subsecuMI}
\begin{figure}
	\centering
	\includegraphics[width=8cm]{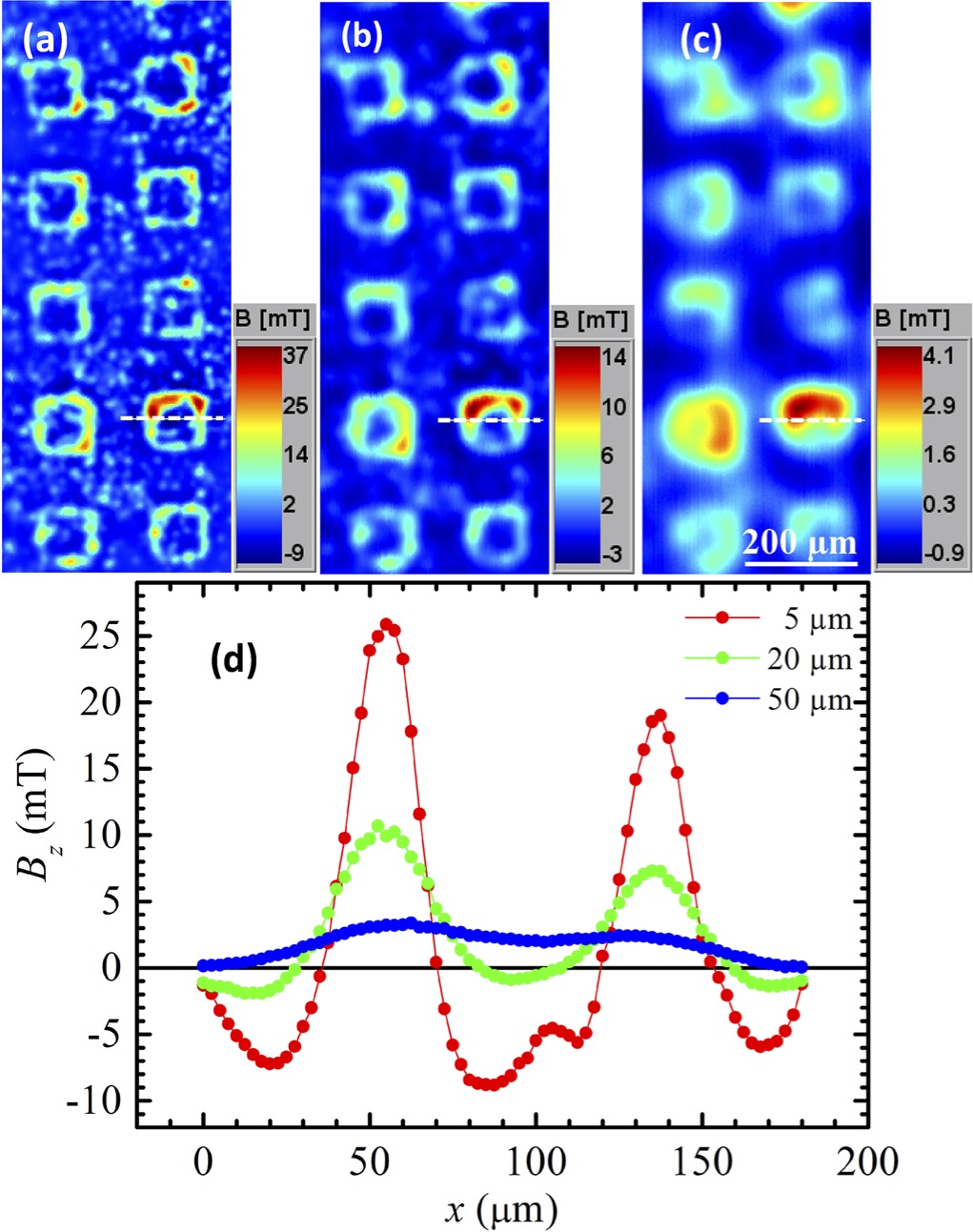}
	\caption{(a)-(c) SHPM images showing the $B_{z}$ distribution above a $\mu$MI structure of an array of 100 $\mu$m $\times$ 100 $\mu$m squares of NdFeB powder at height $h$ = 5, 10 and 20 $\mu$m, respectively. The individual color bars are shown alongside each image. (d) $B_{z}(x)$ profiles across the dashed lines shown on each image.}
	\label{f7}
\end{figure}

We have characterized hard magnetic powder (NdFeB) based micro-flux sources prepared using the micro-magnetic imprinting ($\mu$µMI) technique \cite{Dempsey2014}. More details on the $\mu$MI technique and preparation and characterization of micro-flux sources by this method can be found in Ref. \citenum{Dempsey2014}. The images in Figures \ref{f7}(a)-(c) show the $z$-component of the stray field distribution $(B_{z})$ produced at heights $h$ = (a) 5 $\mu$m, (b) 20 $\mu$m, and (c) 50 $\mu$m ($\pm$ 1 $\mu$m) above a $\mu$MI structure of an array of 100 $\mu$m $\times$ 100 $\mu$m squares of NdFeB powder (average NdFeB particle size = 5 $\mu$m). For each image, the scan resolution = 2.5 $\mu$m and the scan area = 160 $\times$ 400 pixel = 400 $\mu$m $\times$ 1000 $\mu$m. Figure \ref{f7}(d) shows $B_{z}(x)$ profiles across the dashed lines shown on each image. Close to the sample surface, the SHPM technique is able to clearly resolve the well-defined individual square structures and large stray field values are detected (cf. image in Figure \ref{f7}(a) at $h$ = 5 $\mu$m and peak-to-peak value of $\sim$ 40 mT in the corresponding $B_{z}(x)$ profile in Figure \ref{f7}(d)). Further, smaller agglomerations of powders across the sample surface are also clearly observed in the image. With increasing scan height, as expected, the individual structures become less resolved and the detected stray fields also decrease. The measurements have been shown to agree very well with simulations of the $z$-component of the stray magnetic field produced above such structures made with isotropic hard magnetic spheres \cite{Dempsey2014}.

\begin{figure}
	\centering
	\includegraphics[width=6 cm]{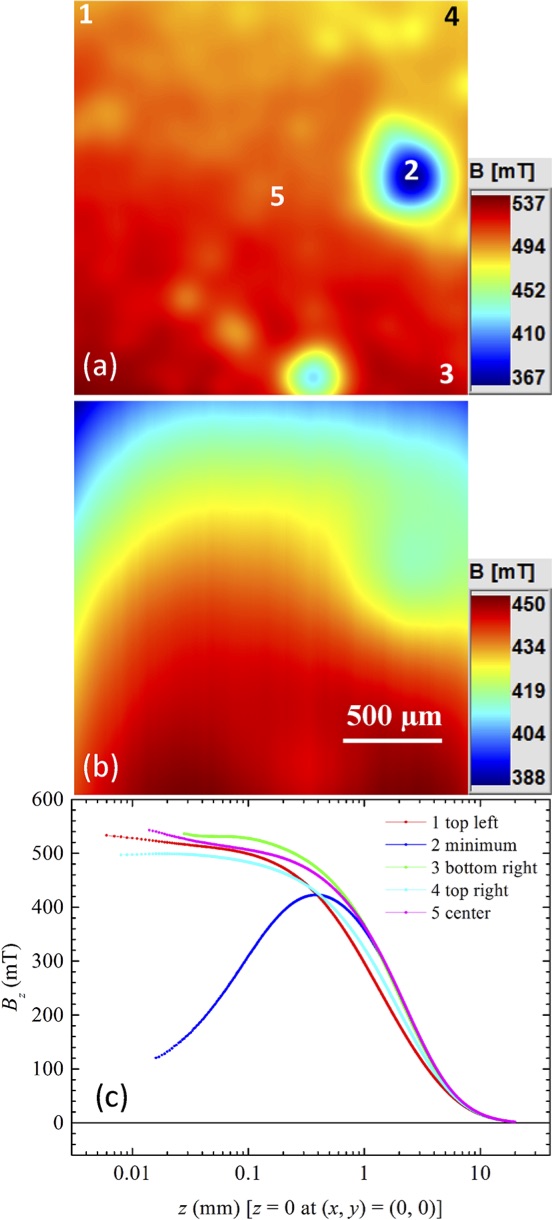}
	\caption{(a) and (b) SHPM images showing the $B_{z}$ distribution above a 5 mm cubic bulk magnet at scan height 100 and 500 $\mu$m, respectively. The individual color bars are shown alongside each image. (c) $B_{z}(z)$ profiles at different locations on the image as indicated by numbers in (a).}
	\label{f8}
\end{figure}

\subsection{Bulk magnets}
\label{subsecBulkMagnets}
A SHPM system is a potentially attractive tool for characterization of bulk magnets, as on one hand it can detect large magnetic fields, and on the other hand, it can reveal local inhomogeneities that could affect the magnet performance. Our setup has been utilized for characterization of a number of mm sized bulk magnets. Figure \ref{f8} shows a summary of measurements on a cubic magnet (5 mm side length). SHPM images in Figures \ref{f8}(a) and \ref{f8}(b) show the $B_{z}$ distribution across a 2 mm $\times$ 2 mm area centered near the center of the magnet, at scan height $h$ = 100 $\mu$m and 500 $\mu$m ($\pm$ 1 $\mu$m), respectively. Note that these mentioned heights are the heights of the probe from the sample surface at the location $(x, y) = (0, 0)$, i.e., the top-left of the image. The actual height at each point would vary according to the roughness of the magnet surface as in these sets the scans are performed in smooth horizontal planes (in flyover mode, using an artificial flat topography for reference). This ensures surface corrugations do not affect the measured $B_{z}$ profiles. The scan resolution is 2.5 $\mu$m (image area 800 $\times$ 800 pixel). Figure \ref{f8}(a) demonstrates inhomogeneities in the magnetic field distribution across the surface of the bulk magnet. The $B_{z}$ value is seen to vary in the large range of 360 mT $< B_{z} <$ 540 mT over the image area. Moreover, pockets of low $B_{z}$ (e.g., cf. dark blue region labelled 2 in Figure \ref{f8}(a)) are observed. These observations are useful because these bulk magnets are used for a variety of purposes, e.g., to magnetize the soft magnetic microstructures discussed in Section \ref{subsecFeCoPillars}. In view of the inhomogeneous $B_{z}$ distribution over the magnet surface, when such bulk magnets are used to characterize flux concentration in soft micro-magnets, it becomes important to take care in choosing the region of the surface onto which the micro-magnet array is placed. However, the image in Figure \ref{f8}(b) at a larger height of 500 $\mu$m from the magnet surface shows that the $B_{z}$ distribution is largely smoothened out at this height from the magnet surface. This means at these heights, the issue of magnetic field inhomogeneity is not as crucial as in the earlier case. 

This is further clarified from local $B_{z}(z)$ profiles at different locations on the magnet. Figure \ref{f8}(c) shows $B_{z}(z)$ profiles obtained at five different locations on the image as indicated by the numbers in Figure \ref{f8}(a). Interestingly, while the profiles at the corners (locations 1, 3 and 4) and center of the image (location 5) show expected monotonic decrease in $B_{z}$ with increasing $z$, in the profile at the center of the low $B_{z}$ region (location 2), the $B_{z}$ value initially increases with increasing $z$. This continues till $z \sim$ 0.4 mm, beyond which the curve reverses and follows the same trend as the others. With increasing $z$, the difference in local $B_{z}$ values at different locations decreases. This is because the detected \textquoteleft local\textquoteright\space field really is an average of field lines emanating from the area surrounding a given location. With increasing $z$, the area over which the averaging occurs increases, and hence local \textquoteleft anomalies\textquoteright\space are averaged out. In general at a distance of $\sim$ 1-2 mm the $B_{z}$ distribution becomes homogeneous over an image size corresponding to the height, though there is still noticeable difference between the local $B_{z}$ values even at heights as large as 5 mm (cf. Figure \ref{f8}). Beyond $z \sim$ 10 mm, all the curves merge into a single one, indicating uniform $B_{z}(z)$ across the magnet surface over the 2 mm $\times$ 2 mm area at these heights. Note that in both images in Figures \ref{f8}(a) and \ref{f8}(b), there is an overall gradient in the $B_{z}$ distribution along the $y$-direction (top-to-bottom of the images). This is attributed to the fact that the Hall probe is kept at a tilt of $\alpha \sim$ 10$^{\circ}$ w.r.t. the sample surface, the angle α being in the $yz$-plane. Hence the effective measured component of $B_{z}$ changes as the probe is scanned along the $y$-axis. Another factor contributing to this gradient could be that the scan area of 2 $\times$ 2 mm was not exactly centered at the middle of the magnet. Note that there is no such gradient along the $x$-axis. This issue is particularly noticeable in this case due to the intense fields and large area of the magnet. In usual scenarios of smaller sized micro-magnet samples this can be ignored, as seen in Sections \ref{subsecFeCoPillars} and \ref{subsecuMI}.

\section{Conclusion}
\label{secConclusion}
We have developed a Scanning Hall Probe Microscope capable of performing high resolution scans (step size of 0.1 $\mu$m) over large areas (few mm, limited by sample topography and tilt). New generation of mesa etched and precision diced Hall sensors enable approaching close to the sample surface ($<$ 2 $\mu$m) resulting in images with highly resolved features (spatial resolution of magnetic field distribution limited by the size of Hall probes, 1 $\mu$m). These Hall probes allow detection of large magnetic fields ($\sim$ 1 T) with high field resolution (100 $\mu$T, limited by system noise during scanning). The AFM-like tuning fork based force detection technique enables imaging in regulation mode (probe in contact with the sample surface) and the microscope can be also used for imaging in flyover mode at variable heights from the sample surface (up to 35 mm). Apart from scanning, the system is capable of recording local $B(z)$ profiles. A single instrument (the Zurich Instruments HF2LI Lockin amplifier) serves as the primary electronics component of the microscope which significantly improves system performance. New programs have been developed for reliable operation and several enhanced functionalities such as precise sensor calibration, more efficient scanning in sets, fast acquisition of $B(z)$ profiles etc. The simple design, three large range, precise stepper motors and a simple piezo stack combined with a reliable tuning fork based height control makes this microscope a very versatile tool for the control and inspection of devices and materials. The setup is being extensively used to measure stray field patterns of micro-magnet arrays as well as for characterization of bulk magnets.


\begin{acknowledgments}
This work was partially funded by the ANR \textquotedblleft MIME\textquotedblright project (ANR-11-BSV5-0101). The work of R. B. G. K. is partially supported by the \textquotedblleft Chaire Initiative Universitaire Alpes (IUA)\textquotedblright of the University Grenoble Alpes. The authors would like to thank T. Devillers, D Le Roy, F. Dumas-Bouchiat, H. Marelli-Mathevon, A. Dias, and D. Givord for providing samples and for useful discussions. G. S. would like to thank the electronic workshop at Institut N\'{e}el for technical support and S. Le-Denmat for his inputs. The authors acknowledge the contributions of D. J. Hykel, M. Kustov and P. Laczkowski in development of the earlier generation of the microscope. The 2DEG wafers used for probe fabrication were provided by Picogiga (Soitec). Probe patterning was carried out at the Nanofab cleanroom facility of Institut N\'{e}el.
\end{acknowledgments}

\nocite{*}

\begin{thebibliography}{0}%
\makeatletter
\providecommand \@ifxundefined [1]{%
 \@ifx{#1\undefined}
}%
\providecommand \@ifnum [1]{%
 \ifnum #1\expandafter \@firstoftwo
 \else \expandafter \@secondoftwo
 \fi
}%
\providecommand \@ifx [1]{%
 \ifx #1\expandafter \@firstoftwo
 \else \expandafter \@secondoftwo
 \fi
}%
\providecommand \natexlab [1]{#1}%
\providecommand \enquote  [1]{``#1''}%
\providecommand \bibnamefont  [1]{#1}%
\providecommand \bibfnamefont [1]{#1}%
\providecommand \citenamefont [1]{#1}%
\providecommand \href@noop [0]{\@secondoftwo}%
\providecommand \href [0]{\begingroup \@sanitize@url \@href}%
\providecommand \@href[1]{\@@startlink{#1}\@@href}%
\providecommand \@@href[1]{\endgroup#1\@@endlink}%
\providecommand \@sanitize@url [0]{\catcode `\\12\catcode `\$12\catcode
  `\&12\catcode `\#12\catcode `\^12\catcode `\_12\catcode `\%12\relax}%
\providecommand \@@startlink[1]{}%
\providecommand \@@endlink[0]{}%
\providecommand \url  [0]{\begingroup\@sanitize@url \@url }%
\providecommand \@url [1]{\endgroup\@href {#1}{\urlprefix }}%
\providecommand \urlprefix  [0]{URL }%
\providecommand \Eprint [0]{\href }%
\providecommand \doibase [0]{http://dx.doi.org/}%
\providecommand \selectlanguage [0]{\@gobble}%
\providecommand \bibinfo  [0]{\@secondoftwo}%
\providecommand \bibfield  [0]{\@secondoftwo}%
\providecommand \translation [1]{[#1]}%
\providecommand \BibitemOpen [0]{}%
\providecommand \bibitemStop [0]{}%
\providecommand \bibitemNoStop [0]{.\EOS\space}%
\providecommand \EOS [0]{\spacefactor3000\relax}%
\providecommand \BibitemShut  [1]{\csname bibitem#1\endcsname}%
\let\auto@bib@innerbib\@empty
\end{thebibliography}%


\begin{thebibliography}{35}%
	\makeatletter
	\providecommand \@ifxundefined [1]{%
		\@ifx{#1\undefined}
	}%
	\providecommand \@ifnum [1]{%
		\ifnum #1\expandafter \@firstoftwo
		\else \expandafter \@secondoftwo
		\fi
	}%
	\providecommand \@ifx [1]{%
		\ifx #1\expandafter \@firstoftwo
		\else \expandafter \@secondoftwo
		\fi
	}%
	\providecommand \natexlab [1]{#1}%
	\providecommand \enquote  [1]{``#1''}%
	\providecommand \bibnamefont  [1]{#1}%
	\providecommand \bibfnamefont [1]{#1}%
	\providecommand \citenamefont [1]{#1}%
	\providecommand \href@noop [0]{\@secondoftwo}%
	\providecommand \href [0]{\begingroup \@sanitize@url \@href}%
	\providecommand \@href[1]{\@@startlink{#1}\@@href}%
	\providecommand \@@href[1]{\endgroup#1\@@endlink}%
	\providecommand \@sanitize@url [0]{\catcode `\\12\catcode `\$12\catcode
		`\&12\catcode `\#12\catcode `\^12\catcode `\_12\catcode `\%12\relax}%
	\providecommand \@@startlink[1]{}%
	\providecommand \@@endlink[0]{}%
	\providecommand \url  [0]{\begingroup\@sanitize@url \@url }%
	\providecommand \@url [1]{\endgroup\@href {#1}{\urlprefix }}%
	\providecommand \urlprefix  [0]{URL }%
	\providecommand \Eprint [0]{\href }%
	\providecommand \doibase [0]{http://dx.doi.org/}%
	\providecommand \selectlanguage [0]{\@gobble}%
	\providecommand \bibinfo  [0]{\@secondoftwo}%
	\providecommand \bibfield  [0]{\@secondoftwo}%
	\providecommand \translation [1]{[#1]}%
	\providecommand \BibitemOpen [0]{}%
	\providecommand \bibitemStop [0]{}%
	\providecommand \bibitemNoStop [0]{.\EOS\space}%
	\providecommand \EOS [0]{\spacefactor3000\relax}%
	\providecommand \BibitemShut  [1]{\csname bibitem#1\endcsname}%
	\let\auto@bib@innerbib\@empty
	\bibitem [{\citenamefont {Chang}\ \emph {et~al.}(1992)\citenamefont {Chang},
		\citenamefont {Hallen}, \citenamefont {Harriott}, \citenamefont {Hess},
		\citenamefont {Kao}, \citenamefont {Kwo}, \citenamefont {Miller},
		\citenamefont {Wolfe}, \citenamefont {van~der Ziel},\ and\ \citenamefont
		{Chang}}]{Chang1992}%
	\BibitemOpen
	\bibfield  {author} {\bibinfo {author} {\bibfnamefont {A.~M.}\ \bibnamefont
			{Chang}}, \bibinfo {author} {\bibfnamefont {H.~D.}\ \bibnamefont {Hallen}},
		\bibinfo {author} {\bibfnamefont {L.}~\bibnamefont {Harriott}}, \bibinfo
		{author} {\bibfnamefont {H.~F.}\ \bibnamefont {Hess}}, \bibinfo {author}
		{\bibfnamefont {H.~L.}\ \bibnamefont {Kao}}, \bibinfo {author} {\bibfnamefont
			{J.}~\bibnamefont {Kwo}}, \bibinfo {author} {\bibfnamefont {R.~E.}\
			\bibnamefont {Miller}}, \bibinfo {author} {\bibfnamefont {R.}~\bibnamefont
			{Wolfe}}, \bibinfo {author} {\bibfnamefont {J.}~\bibnamefont {van~der Ziel}},
		\ and\ \bibinfo {author} {\bibfnamefont {T.~Y.}\ \bibnamefont {Chang}},\
	}\bibfield  {title} {\enquote {\bibinfo {title} {Scanning {Hall} probe
				microscopy},}\ }\href {\doibase 10.1063/1.108334} {\bibfield  {journal}
		{\bibinfo  {journal} {Appl. Phys. Lett.}\ }\textbf {\bibinfo {volume} {61}},\
		\bibinfo {pages} {1974} (\bibinfo {year} {1992})}\BibitemShut {NoStop}%
	\bibitem [{\citenamefont {Oral}, \citenamefont {Bending},\ and\ \citenamefont
		{Henini}(1996{\natexlab{a}})}]{Oral1996}%
	\BibitemOpen
	\bibfield  {author} {\bibinfo {author} {\bibfnamefont {A.}~\bibnamefont
			{Oral}}, \bibinfo {author} {\bibfnamefont {S.~J.}\ \bibnamefont {Bending}}, \
		and\ \bibinfo {author} {\bibfnamefont {M.}~\bibnamefont {Henini}},\
	}\bibfield  {title} {\enquote {\bibinfo {title} {Real-time scanning hall
				probe microscopy},}\ }\href {\doibase 10.1063/1.117582} {\bibfield  {journal}
		{\bibinfo  {journal} {Applied Physics Letters}\ }\textbf {\bibinfo {volume}
			{69}},\ \bibinfo {pages} {1324} (\bibinfo {year}
		{1996}{\natexlab{a}})}\BibitemShut {NoStop}%
	\bibitem [{\citenamefont {Oral}, \citenamefont {Bending},\ and\ \citenamefont
		{Henini}(1996{\natexlab{b}})}]{Oral1996a}%
	\BibitemOpen
	\bibfield  {author} {\bibinfo {author} {\bibfnamefont {A.}~\bibnamefont
			{Oral}}, \bibinfo {author} {\bibfnamefont {S.~J.}\ \bibnamefont {Bending}}, \
		and\ \bibinfo {author} {\bibfnamefont {M.}~\bibnamefont {Henini}},\
	}\bibfield  {title} {\enquote {\bibinfo {title} {Scanning {Hall} probe
				microscopy of superconductors and magnetic materials},}\ }\href {\doibase
		10.1116/1.588514} {\bibfield  {journal} {\bibinfo  {journal} {Journal of
				Vacuum Science \& Technology B}\ }\textbf {\bibinfo {volume} {14}},\ \bibinfo
		{pages} {1202--1205} (\bibinfo {year} {1996}{\natexlab{b}})}\BibitemShut
	{NoStop}%
	\bibitem [{\citenamefont {Howells}\ \emph {et~al.}(1999)\citenamefont
		{Howells}, \citenamefont {Oral}, \citenamefont {Bending}, \citenamefont
		{Andrews}, \citenamefont {Squire}, \citenamefont {Rice}, \citenamefont
		{de~Lozanne}, \citenamefont {Bland}, \citenamefont {Kaya},\ and\
		\citenamefont {Henini}}]{Howells1999}%
	\BibitemOpen
	\bibfield  {author} {\bibinfo {author} {\bibfnamefont {G.}~\bibnamefont
			{Howells}}, \bibinfo {author} {\bibfnamefont {A.}~\bibnamefont {Oral}},
		\bibinfo {author} {\bibfnamefont {S.}~\bibnamefont {Bending}}, \bibinfo
		{author} {\bibfnamefont {S.}~\bibnamefont {Andrews}}, \bibinfo {author}
		{\bibfnamefont {P.}~\bibnamefont {Squire}}, \bibinfo {author} {\bibfnamefont
			{P.}~\bibnamefont {Rice}}, \bibinfo {author} {\bibfnamefont {A.}~\bibnamefont
			{de~Lozanne}}, \bibinfo {author} {\bibfnamefont {J.}~\bibnamefont {Bland}},
		\bibinfo {author} {\bibfnamefont {I.}~\bibnamefont {Kaya}}, \ and\ \bibinfo
		{author} {\bibfnamefont {M.}~\bibnamefont {Henini}},\ }\bibfield  {title}
	{\enquote {\bibinfo {title} {Scanning {Hall} probe microscopy of
				ferromagnetic structures},}\ }\href {\doibase 10.1016/s0304-8853(98)01002-6}
	{\bibfield  {journal} {\bibinfo  {journal} {J. Magn. Magn. Mater.}\ }\textbf
		{\bibinfo {volume} {196-197}},\ \bibinfo {pages} {917â€“919} (\bibinfo
		{year} {1999})}\BibitemShut {NoStop}%
	\bibitem [{\citenamefont {Sandhu}\ \emph {et~al.}(2002)\citenamefont {Sandhu},
		\citenamefont {Masuda}, \citenamefont {Oral}, \citenamefont {Bending},
		\citenamefont {Yamada},\ and\ \citenamefont {Konagai}}]{Sandhu2002}%
	\BibitemOpen
	\bibfield  {author} {\bibinfo {author} {\bibfnamefont {A.}~\bibnamefont
			{Sandhu}}, \bibinfo {author} {\bibfnamefont {H.}~\bibnamefont {Masuda}},
		\bibinfo {author} {\bibfnamefont {A.}~\bibnamefont {Oral}}, \bibinfo {author}
		{\bibfnamefont {S.}~\bibnamefont {Bending}}, \bibinfo {author} {\bibfnamefont
			{A.}~\bibnamefont {Yamada}}, \ and\ \bibinfo {author} {\bibfnamefont
			{M.}~\bibnamefont {Konagai}},\ }\bibfield  {title} {\enquote {\bibinfo
			{title} {Room temperature scanning {Hall} probe microscopy using
				{GaAs}/{AlGaAs} and bi micro-hall probes},}\ }\href {\doibase
		10.1016/s0304-3991(02)00087-6} {\bibfield  {journal} {\bibinfo  {journal}
			{Ultramicroscopy}\ }\textbf {\bibinfo {volume} {91}},\ \bibinfo {pages}
		{97â€“101} (\bibinfo {year} {2002})}\BibitemShut {NoStop}%
	\bibitem [{\citenamefont {Bending}(2010)}]{Bending2010}%
	\BibitemOpen
	\bibfield  {author} {\bibinfo {author} {\bibfnamefont {S.~J.}\ \bibnamefont
			{Bending}},\ }\bibfield  {title} {\enquote {\bibinfo {title} {Scanning {Hall}
				probe microscopy of vortex matter},}\ }\href {\doibase
		10.1016/j.physc.2010.02.027} {\bibfield  {journal} {\bibinfo  {journal}
			{Physica C}\ }\textbf {\bibinfo {volume} {470}},\ \bibinfo {pages}
		{754â€“757} (\bibinfo {year} {2010})}\BibitemShut {NoStop}%
	\bibitem [{\citenamefont {Kirtley}(2010)}]{Kirtley2010}%
	\BibitemOpen
	\bibfield  {author} {\bibinfo {author} {\bibfnamefont {J.~R.}\ \bibnamefont
			{Kirtley}},\ }\bibfield  {title} {\enquote {\bibinfo {title} {Fundamental
				studies of superconductors using scanning magnetic imaging},}\ }\href
	{\doibase 10.1088/0034-4885/73/12/126501} {\bibfield  {journal} {\bibinfo
			{journal} {Rep. Prog. Phys.}\ }\textbf {\bibinfo {volume} {73}},\ \bibinfo
		{pages} {126501} (\bibinfo {year} {2010})}\BibitemShut {NoStop}%
	\bibitem [{\citenamefont {Gregory}, \citenamefont {Bending},\ and\
		\citenamefont {Sandhu}(2002)}]{Gregory2002}%
	\BibitemOpen
	\bibfield  {author} {\bibinfo {author} {\bibfnamefont {J.~K.}\ \bibnamefont
			{Gregory}}, \bibinfo {author} {\bibfnamefont {S.~J.}\ \bibnamefont
			{Bending}}, \ and\ \bibinfo {author} {\bibfnamefont {A.}~\bibnamefont
			{Sandhu}},\ }\bibfield  {title} {\enquote {\bibinfo {title} {A scanning
				{Hall} probe microscope for large area magnetic imaging down to cryogenic
				temperatures},}\ }\href {\doibase 10.1063/1.1505097} {\bibfield  {journal}
		{\bibinfo  {journal} {Rev. Sci. Instrum.}\ }\textbf {\bibinfo {volume}
			{73}},\ \bibinfo {pages} {3515} (\bibinfo {year} {2002})}\BibitemShut
	{NoStop}%
	\bibitem [{\citenamefont {Dinner}, \citenamefont {Beasley},\ and\ \citenamefont
		{Moler}(2005)}]{Dinner2005}%
	\BibitemOpen
	\bibfield  {author} {\bibinfo {author} {\bibfnamefont {R.~B.}\ \bibnamefont
			{Dinner}}, \bibinfo {author} {\bibfnamefont {M.~R.}\ \bibnamefont {Beasley}},
		\ and\ \bibinfo {author} {\bibfnamefont {K.~A.}\ \bibnamefont {Moler}},\
	}\bibfield  {title} {\enquote {\bibinfo {title} {Cryogenic scanning
				{Hall-probe} microscope with centimeter scan range and submicron
				resolution},}\ }\href {\doibase 10.1063/1.2072438} {\bibfield  {journal}
		{\bibinfo  {journal} {Rev. Sci. Instrum.}\ }\textbf {\bibinfo {volume}
			{76}},\ \bibinfo {pages} {103702} (\bibinfo {year} {2005})}\BibitemShut
	{NoStop}%
	\bibitem [{\citenamefont {Cambel}\ \emph {et~al.}(2005)\citenamefont {Cambel},
		\citenamefont {Fedor}, \citenamefont {GreguÅ¡ovÃ¡}, \citenamefont
		{KovÃ¡Ä�},\ and\ \citenamefont {HuÅ¡ek}}]{Cambel2005}%
	\BibitemOpen
	\bibfield  {author} {\bibinfo {author} {\bibfnamefont {V.}~\bibnamefont
			{Cambel}}, \bibinfo {author} {\bibfnamefont {J.}~\bibnamefont {Fedor}},
		\bibinfo {author} {\bibfnamefont {D.}~\bibnamefont {GreguÅ¡ovÃ¡}},
		\bibinfo {author} {\bibfnamefont {P.}~\bibnamefont {KovÃ¡Ä�}}, \ and\
		\bibinfo {author} {\bibfnamefont {I.}~\bibnamefont {HuÅ¡ek}},\ }\bibfield
	{title} {\enquote {\bibinfo {title} {Large-scale high-resolution scanning
				hall probe microscope used for mgb 2 filament characterization},}\ }\href
	{\doibase 10.1088/0953-2048/18/4/007} {\bibfield  {journal} {\bibinfo
			{journal} {Supercond. Sci. Technol.}\ }\textbf {\bibinfo {volume} {18}},\
		\bibinfo {pages} {417â€“421} (\bibinfo {year} {2005})}\BibitemShut
	{NoStop}%
	\bibitem [{\citenamefont {Tang}\ \emph {et~al.}(2014)\citenamefont {Tang},
		\citenamefont {Lin}, \citenamefont {Wu}, \citenamefont {Chen}, \citenamefont
		{Wang}, \citenamefont {Ling}, \citenamefont {Chi},\ and\ \citenamefont
		{Chen}}]{Tang2014}%
	\BibitemOpen
	\bibfield  {author} {\bibinfo {author} {\bibfnamefont {C.-C.}\ \bibnamefont
			{Tang}}, \bibinfo {author} {\bibfnamefont {H.-T.}\ \bibnamefont {Lin}},
		\bibinfo {author} {\bibfnamefont {S.-L.}\ \bibnamefont {Wu}}, \bibinfo
		{author} {\bibfnamefont {T.-J.}\ \bibnamefont {Chen}}, \bibinfo {author}
		{\bibfnamefont {M.~J.}\ \bibnamefont {Wang}}, \bibinfo {author}
		{\bibfnamefont {D.~C.}\ \bibnamefont {Ling}}, \bibinfo {author}
		{\bibfnamefont {C.~C.}\ \bibnamefont {Chi}}, \ and\ \bibinfo {author}
		{\bibfnamefont {J.-C.}\ \bibnamefont {Chen}},\ }\bibfield  {title} {\enquote
		{\bibinfo {title} {An interchangeable scanning hall probe/scanning squid
				microscope},}\ }\href {\doibase 10.1063/1.4893647} {\bibfield  {journal}
		{\bibinfo  {journal} {Review of Scientific Instruments}\ }\textbf {\bibinfo
			{volume} {85}},\ \bibinfo {pages} {083707} (\bibinfo {year}
		{2014})}\BibitemShut {NoStop}%
	\bibitem [{\citenamefont {\"{O}zg\"{u}r Karc{\i}}\ \emph
		{et~al.}(2014)\citenamefont {\"{O}zg\"{u}r Karc{\i}}, \citenamefont {Piatek},
		\citenamefont {Jorba}, \citenamefont {Dede}, \citenamefont {R{\o}nnow},\ and\
		\citenamefont {Oral}}]{Karci2014}%
	\BibitemOpen
	\bibfield  {author} {\bibinfo {author} {\bibnamefont {\"{O}zg\"{u}r
				Karc{\i}}}, \bibinfo {author} {\bibfnamefont {J.~O.}\ \bibnamefont {Piatek}},
		\bibinfo {author} {\bibfnamefont {P.}~\bibnamefont {Jorba}}, \bibinfo
		{author} {\bibfnamefont {M.}~\bibnamefont {Dede}}, \bibinfo {author}
		{\bibfnamefont {H.~M.}\ \bibnamefont {R{\o}nnow}}, \ and\ \bibinfo {author}
		{\bibfnamefont {A.}~\bibnamefont {Oral}},\ }\bibfield  {title} {\enquote
		{\bibinfo {title} {{An ultra-low temperature scanning Hall probe microscope
					for magnetic imaging below 40 {mK}}},}\ }\href {\doibase 10.1063/1.4897145}
	{\bibfield  {journal} {\bibinfo  {journal} {Rev. Sci. Instrum.}\ }\textbf
		{\bibinfo {volume} {85}},\ \bibinfo {pages} {103703} (\bibinfo {year}
		{2014})}\BibitemShut {NoStop}%
	\bibitem [{\citenamefont {Shimizu}\ \emph {et~al.}(2004)\citenamefont
		{Shimizu}, \citenamefont {Saitoh}, \citenamefont {Miyajima},\ and\
		\citenamefont {Masuda}}]{Shimizu2004}%
	\BibitemOpen
	\bibfield  {author} {\bibinfo {author} {\bibfnamefont {M.}~\bibnamefont
			{Shimizu}}, \bibinfo {author} {\bibfnamefont {E.}~\bibnamefont {Saitoh}},
		\bibinfo {author} {\bibfnamefont {H.}~\bibnamefont {Miyajima}}, \ and\
		\bibinfo {author} {\bibfnamefont {H.}~\bibnamefont {Masuda}},\ }\bibfield
	{title} {\enquote {\bibinfo {title} {Scanning {Hall} probe microscopy with
				high resolution of magnetic field image},}\ }\href {\doibase
		10.1016/j.jmmm.2004.04.086} {\bibfield  {journal} {\bibinfo  {journal} {J.
				Magn. Magn. Mater.}\ }\textbf {\bibinfo {volume} {282}},\ \bibinfo {pages}
		{369â€“372} (\bibinfo {year} {2004})}\BibitemShut {NoStop}%
	\bibitem [{\citenamefont {Zanini}\ \emph {et~al.}(2011)\citenamefont {Zanini},
		\citenamefont {Dempsey}, \citenamefont {Givord}, \citenamefont {Reyne},\ and\
		\citenamefont {Dumas-Bouchiat}}]{Zanini2011}%
	\BibitemOpen
	\bibfield  {author} {\bibinfo {author} {\bibfnamefont {L.~F.}\ \bibnamefont
			{Zanini}}, \bibinfo {author} {\bibfnamefont {N.~M.}\ \bibnamefont {Dempsey}},
		\bibinfo {author} {\bibfnamefont {D.}~\bibnamefont {Givord}}, \bibinfo
		{author} {\bibfnamefont {G.}~\bibnamefont {Reyne}}, \ and\ \bibinfo {author}
		{\bibfnamefont {F.}~\bibnamefont {Dumas-Bouchiat}},\ }\bibfield  {title}
	{\enquote {\bibinfo {title} {Autonomous micro-magnet based systems for highly
				efficient magnetic separation},}\ }\href {\doibase 10.1063/1.3664092}
	{\bibfield  {journal} {\bibinfo  {journal} {Appl. Phys. Lett.}\ }\textbf
		{\bibinfo {volume} {99}},\ \bibinfo {pages} {232504} (\bibinfo {year}
		{2011})}\BibitemShut {NoStop}%
	\bibitem [{\citenamefont {Brunet}\ \emph {et~al.}(2013)\citenamefont {Brunet},
		\citenamefont {Bouclet}, \citenamefont {Ahmadi}, \citenamefont {Mitrossilis},
		\citenamefont {Driquez}, \citenamefont {Brunet}, \citenamefont {Henry},
		\citenamefont {Serman}, \citenamefont {B{\'{e}}alle}, \citenamefont
		{M{\'{e}}nager}, \citenamefont {Dumas-Bouchiat}, \citenamefont {Givord},
		\citenamefont {Yanicostas}, \citenamefont {Le-Roy}, \citenamefont {Dempsey},
		\citenamefont {Plessis},\ and\ \citenamefont {Farge}}]{Brunet2013}%
	\BibitemOpen
	\bibfield  {author} {\bibinfo {author} {\bibfnamefont {T.}~\bibnamefont
			{Brunet}}, \bibinfo {author} {\bibfnamefont {A.}~\bibnamefont {Bouclet}},
		\bibinfo {author} {\bibfnamefont {P.}~\bibnamefont {Ahmadi}}, \bibinfo
		{author} {\bibfnamefont {D.}~\bibnamefont {Mitrossilis}}, \bibinfo {author}
		{\bibfnamefont {B.}~\bibnamefont {Driquez}}, \bibinfo {author} {\bibfnamefont
			{A.-C.}\ \bibnamefont {Brunet}}, \bibinfo {author} {\bibfnamefont
			{L.}~\bibnamefont {Henry}}, \bibinfo {author} {\bibfnamefont
			{F.}~\bibnamefont {Serman}}, \bibinfo {author} {\bibfnamefont
			{G.}~\bibnamefont {B{\'{e}}alle}}, \bibinfo {author} {\bibfnamefont
			{C.}~\bibnamefont {M{\'{e}}nager}}, \bibinfo {author} {\bibfnamefont
			{F.}~\bibnamefont {Dumas-Bouchiat}}, \bibinfo {author} {\bibfnamefont
			{D.}~\bibnamefont {Givord}}, \bibinfo {author} {\bibfnamefont
			{C.}~\bibnamefont {Yanicostas}}, \bibinfo {author} {\bibfnamefont
			{D.}~\bibnamefont {Le-Roy}}, \bibinfo {author} {\bibfnamefont {N.~M.}\
			\bibnamefont {Dempsey}}, \bibinfo {author} {\bibfnamefont {A.}~\bibnamefont
			{Plessis}}, \ and\ \bibinfo {author} {\bibfnamefont {E.}~\bibnamefont
			{Farge}},\ }\bibfield  {title} {\enquote {\bibinfo {title} {Evolutionary
				conservation of early mesoderm specification by mechanotransduction in
				bilateria},}\ }\href {\doibase 10.1038/ncomms3821} {\bibfield  {journal}
		{\bibinfo  {journal} {Nature Communications}\ }\textbf {\bibinfo {volume}
			{4}} (\bibinfo {year} {2013}),\ 10.1038/ncomms3821}\BibitemShut {NoStop}%
	\bibitem [{\citenamefont {Pivetal}\ \emph {et~al.}(2014)\citenamefont
		{Pivetal}, \citenamefont {Ciuta}, \citenamefont {Frenea-Robin}, \citenamefont
		{Haddour}, \citenamefont {Dempsey}, \citenamefont {Dumas-Bouchiat},\ and\
		\citenamefont {Simonet}}]{Pivetal2014}%
	\BibitemOpen
	\bibfield  {author} {\bibinfo {author} {\bibfnamefont {J.}~\bibnamefont
			{Pivetal}}, \bibinfo {author} {\bibfnamefont {G.}~\bibnamefont {Ciuta}},
		\bibinfo {author} {\bibfnamefont {M.}~\bibnamefont {Frenea-Robin}}, \bibinfo
		{author} {\bibfnamefont {N.}~\bibnamefont {Haddour}}, \bibinfo {author}
		{\bibfnamefont {N.~M.}\ \bibnamefont {Dempsey}}, \bibinfo {author}
		{\bibfnamefont {F.}~\bibnamefont {Dumas-Bouchiat}}, \ and\ \bibinfo {author}
		{\bibfnamefont {P.}~\bibnamefont {Simonet}},\ }\bibfield  {title} {\enquote
		{\bibinfo {title} {Magnetic nanoparticle {DNA} labeling for individual
				bacterial cell detection and recovery},}\ }\href {\doibase
		10.1016/j.mimet.2014.09.006} {\bibfield  {journal} {\bibinfo  {journal}
			{Journal of Microbiological Methods}\ }\textbf {\bibinfo {volume} {107}},\
		\bibinfo {pages} {84--91} (\bibinfo {year} {2014})}\BibitemShut {NoStop}%
	\bibitem [{\citenamefont {Dempsey}(2009)}]{Dempsey2009}%
	\BibitemOpen
	\bibfield  {author} {\bibinfo {author} {\bibfnamefont {N.~M.}\ \bibnamefont
			{Dempsey}},\ }\bibfield  {title} {\enquote {\bibinfo {title} {Hard magnetic
				materials for {MEMS} applications},}\ }in\ \href {\doibase
		10.1007/978-0-387-85600-1_22} {\emph {\bibinfo {booktitle} {Nanoscale
				Magnetic Materials and Applications}}}\ (\bibinfo  {publisher} {Springer
		Science+Business Media},\ \bibinfo {year} {2009})\ pp.\ \bibinfo {pages}
	{661--683}\BibitemShut {NoStop}%
	\bibitem [{\citenamefont {Kustov}\ \emph {et~al.}(2010)\citenamefont {Kustov},
		\citenamefont {Laczkowski}, \citenamefont {Hykel}, \citenamefont
		{Hasselbach}, \citenamefont {Dumas-Bouchiat}, \citenamefont {O'Brien},
		\citenamefont {Kauffmann}, \citenamefont {Grechishkin}, \citenamefont
		{Givord}, \citenamefont {Reyne}, \citenamefont {Cugat},\ and\ \citenamefont
		{Dempsey}}]{Kustov2010}%
	\BibitemOpen
	\bibfield  {author} {\bibinfo {author} {\bibfnamefont {M.}~\bibnamefont
			{Kustov}}, \bibinfo {author} {\bibfnamefont {P.}~\bibnamefont {Laczkowski}},
		\bibinfo {author} {\bibfnamefont {D.}~\bibnamefont {Hykel}}, \bibinfo
		{author} {\bibfnamefont {K.}~\bibnamefont {Hasselbach}}, \bibinfo {author}
		{\bibfnamefont {F.}~\bibnamefont {Dumas-Bouchiat}}, \bibinfo {author}
		{\bibfnamefont {D.}~\bibnamefont {O'Brien}}, \bibinfo {author} {\bibfnamefont
			{P.}~\bibnamefont {Kauffmann}}, \bibinfo {author} {\bibfnamefont
			{R.}~\bibnamefont {Grechishkin}}, \bibinfo {author} {\bibfnamefont
			{D.}~\bibnamefont {Givord}}, \bibinfo {author} {\bibfnamefont
			{G.}~\bibnamefont {Reyne}}, \bibinfo {author} {\bibfnamefont
			{O.}~\bibnamefont {Cugat}}, \ and\ \bibinfo {author} {\bibfnamefont {N.~M.}\
			\bibnamefont {Dempsey}},\ }\bibfield  {title} {\enquote {\bibinfo {title}
			{{Magnetic characterization of micropatterned Nd{\textendash}Fe{\textendash}B
					hard magnetic films using scanning Hall probe microscopy}},}\ }\href
	{\doibase 10.1063/1.3486513} {\bibfield  {journal} {\bibinfo  {journal} {J.
				Appl. Phys.}\ }\textbf {\bibinfo {volume} {108}},\ \bibinfo {pages} {063914}
		(\bibinfo {year} {2010})}\BibitemShut {NoStop}%
	\bibitem [{zhi()}]{zhinst}%
	\BibitemOpen
	\href {https://www.zhinst.com/products/hf2li} {}\bibinfo {howpublished}
	{{\url{https://www.zhinst.com/products/hf2li}}}\BibitemShut {NoStop}%
	\bibitem [{pls()}]{pls85}%
	\BibitemOpen
	\href {http://www.pimicos.com/web2/en/1,4,150,pls85.html} {}\bibinfo
	{howpublished}
	{{\url{http://www.pimicos.com/web2/en/1,4,150,pls85.html}}}\BibitemShut
	{NoStop}%
	\bibitem [{ls6()}]{ls65}%
	\BibitemOpen
	\href {http://www.pimicos.com/web2/en/1,4,160,ls65.html} {}\bibinfo
	{howpublished}
	{{\url{http://www.pimicos.com/web2/en/1,4,160,ls65.html}}}\BibitemShut
	{NoStop}%
	\bibitem [{cor()}]{corvus}%
	\BibitemOpen
	\href {http://www.pimicos.com/web2/en/1,2,051,smc_corvus_pci.html} {}\bibinfo
	{howpublished}
	{{\url{http://www.pimicos.com/web2/en/1,2,051,smc_corvus_pci.html}}}\BibitemShut
	{NoStop}%
	\bibitem [{qua()}]{quartz}%
	\BibitemOpen
	\href {http://www.tfc.co.uk/pdfs/watch_crystal_cylindrical.pdf} {}\bibinfo
	{howpublished}
	{{\url{http://www.tfc.co.uk/pdfs/watch_crystal_cylindrical.pdf}}}\BibitemShut
	{NoStop}%
	\bibitem [{\citenamefont {Veauvy}, \citenamefont {Hasselbach},\ and\
		\citenamefont {Mailly}(2002)}]{Veauvy2002}%
	\BibitemOpen
	\bibfield  {author} {\bibinfo {author} {\bibfnamefont {C.}~\bibnamefont
			{Veauvy}}, \bibinfo {author} {\bibfnamefont {K.}~\bibnamefont {Hasselbach}},
		\ and\ \bibinfo {author} {\bibfnamefont {D.}~\bibnamefont {Mailly}},\
	}\bibfield  {title} {\enquote {\bibinfo {title} {Scanning μ-superconduction
				quantum interference device force microscope},}\ }\href {\doibase
		10.1063/1.1515384} {\bibfield  {journal} {\bibinfo  {journal} {Review of
				Scientific Instruments}\ }\textbf {\bibinfo {volume} {73}},\ \bibinfo {pages}
		{3825--3830} (\bibinfo {year} {2002})}\BibitemShut {NoStop}%
	\bibitem [{\citenamefont {Hykel}\ \emph {et~al.}(2014)\citenamefont {Hykel},
		\citenamefont {Wang}, \citenamefont {Castellazzi}, \citenamefont {Crozes},
		\citenamefont {Shaw}, \citenamefont {Schuster},\ and\ \citenamefont
		{Hasselbach}}]{Hykel2014}%
	\BibitemOpen
	\bibfield  {author} {\bibinfo {author} {\bibfnamefont {D.~J.}\ \bibnamefont
			{Hykel}}, \bibinfo {author} {\bibfnamefont {Z.~S.}\ \bibnamefont {Wang}},
		\bibinfo {author} {\bibfnamefont {P.}~\bibnamefont {Castellazzi}}, \bibinfo
		{author} {\bibfnamefont {T.}~\bibnamefont {Crozes}}, \bibinfo {author}
		{\bibfnamefont {G.}~\bibnamefont {Shaw}}, \bibinfo {author} {\bibfnamefont
			{K.}~\bibnamefont {Schuster}}, \ and\ \bibinfo {author} {\bibfnamefont
			{K.}~\bibnamefont {Hasselbach}},\ }\bibfield  {title} {\enquote {\bibinfo
			{title} {{MicroSQUID Force Microscopy in a Dilution Refrigerator}},}\ }\href
	{\doibase 10.1007/s10909-014-1174-9} {\bibfield  {journal} {\bibinfo
			{journal} {J. Low Temp. Phys.}\ }\textbf {\bibinfo {volume} {175}},\ \bibinfo
		{pages} {861--867} (\bibinfo {year} {2014})}\BibitemShut {NoStop}%
	\bibitem [{zpi()}]{zpiezo}%
	\BibitemOpen
	\href@noop {} {}\bibinfo {howpublished} {{Physik Instrumente GmbH and Co. KG:
			P-885.90}}\BibitemShut {NoStop}%
	\bibitem [{\citenamefont {Karrai}\ and\ \citenamefont
		{Grober}(1995)}]{Karrai1995}%
	\BibitemOpen
	\bibfield  {author} {\bibinfo {author} {\bibfnamefont {K.}~\bibnamefont
			{Karrai}}\ and\ \bibinfo {author} {\bibfnamefont {R.~D.}\ \bibnamefont
			{Grober}},\ }\bibfield  {title} {\enquote {\bibinfo {title} {Piezoelectric
				tip-sample distance control for near field optical microscopes},}\ }\href
	{\doibase 10.1063/1.113340} {\bibfield  {journal} {\bibinfo  {journal}
			{Applied Physics Letters}\ }\textbf {\bibinfo {volume} {66}},\ \bibinfo
		{pages} {1842} (\bibinfo {year} {1995})}\BibitemShut {NoStop}%
	\bibitem [{\citenamefont {Karra\"{\i}}\ and\ \citenamefont
		{Grober}(1995)}]{Karrai1995a}%
	\BibitemOpen
	\bibfield  {author} {\bibinfo {author} {\bibfnamefont {K.}~\bibnamefont
			{Karra\"{\i}}}\ and\ \bibinfo {author} {\bibfnamefont {R.~D.}\ \bibnamefont
			{Grober}},\ }\bibfield  {title} {\enquote {\bibinfo {title} {Piezo-electric
				tuning fork tip{\textemdash}sample distance control for near field optical
				microscopes},}\ }\href {\doibase 10.1016/0304-3991(95)00104-2} {\bibfield
		{journal} {\bibinfo  {journal} {Ultramicroscopy}\ }\textbf {\bibinfo {volume}
			{61}},\ \bibinfo {pages} {197--205} (\bibinfo {year} {1995})}\BibitemShut
	{NoStop}%
	\bibitem [{\citenamefont {Dede}\ \emph {et~al.}(2008)\citenamefont {Dede},
		\citenamefont {\"{U}rkmen}, \citenamefont {Giri\c{s}en}, \citenamefont
		{Atabak}, \citenamefont {Oral}, \citenamefont {Farrer},\ and\ \citenamefont
		{Ritchie}}]{Dede2008}%
	\BibitemOpen
	\bibfield  {author} {\bibinfo {author} {\bibfnamefont {M.}~\bibnamefont
			{Dede}}, \bibinfo {author} {\bibfnamefont {K.}~\bibnamefont {\"{U}rkmen}},
		\bibinfo {author} {\bibfnamefont {O.}~\bibnamefont {Giri\c{s}en}}, \bibinfo
		{author} {\bibfnamefont {M.}~\bibnamefont {Atabak}}, \bibinfo {author}
		{\bibfnamefont {A.}~\bibnamefont {Oral}}, \bibinfo {author} {\bibfnamefont
			{I.}~\bibnamefont {Farrer}}, \ and\ \bibinfo {author} {\bibfnamefont
			{D.}~\bibnamefont {Ritchie}},\ }\bibfield  {title} {\enquote {\bibinfo
			{title} {{Scanning Hall Probe Microscopy (SHPM) Using Quartz Crystal AFM
					Feedback}},}\ }\href {\doibase 10.1166/jnn.2008.a265} {\bibfield  {journal}
		{\bibinfo  {journal} {J. Nanosci. Nanotechnol.}\ }\textbf {\bibinfo {volume}
			{8}},\ \bibinfo {pages} {619â€“622} (\bibinfo {year}
		{2008})}\BibitemShut {NoStop}%
	\bibitem [{\citenamefont {Hicks}\ \emph {et~al.}(2007)\citenamefont {Hicks},
		\citenamefont {Luan}, \citenamefont {Moler}, \citenamefont {Zeldov},\ and\
		\citenamefont {Shtrikman}}]{Hicks2007}%
	\BibitemOpen
	\bibfield  {author} {\bibinfo {author} {\bibfnamefont {C.~W.}\ \bibnamefont
			{Hicks}}, \bibinfo {author} {\bibfnamefont {L.}~\bibnamefont {Luan}},
		\bibinfo {author} {\bibfnamefont {K.~A.}\ \bibnamefont {Moler}}, \bibinfo
		{author} {\bibfnamefont {E.}~\bibnamefont {Zeldov}}, \ and\ \bibinfo {author}
		{\bibfnamefont {H.}~\bibnamefont {Shtrikman}},\ }\bibfield  {title} {\enquote
		{\bibinfo {title} {Noise characteristics of 100~{nm} scale {GaAs}∕al[sub
				x]ga[sub 1-x]as scanning hall probes},}\ }\href {\doibase 10.1063/1.2717565}
	{\bibfield  {journal} {\bibinfo  {journal} {Appl. Phys. Lett.}\ }\textbf
		{\bibinfo {volume} {90}},\ \bibinfo {pages} {133512} (\bibinfo {year}
		{2007})}\BibitemShut {NoStop}%
	\bibitem [{\citenamefont {Kustov}(2011)}]{KustovThesis}%
	\BibitemOpen
	\bibfield  {author} {\bibinfo {author} {\bibfnamefont {M.}~\bibnamefont
			{Kustov}},\ }\href@noop {} {Ph.D. thesis},\ \bibinfo  {school} {Université
		Joseph Fourier Grenoble 1} (\bibinfo {year} {2011})\BibitemShut {NoStop}%
	\bibitem [{\citenamefont {Geim}\ \emph {et~al.}(1997)\citenamefont {Geim},
		\citenamefont {Dubonos}, \citenamefont {Lok}, \citenamefont {Grigorieva},
		\citenamefont {Maan}, \citenamefont {Hansen},\ and\ \citenamefont
		{Lindelof}}]{Geim1997}%
	\BibitemOpen
	\bibfield  {author} {\bibinfo {author} {\bibfnamefont {A.~K.}\ \bibnamefont
			{Geim}}, \bibinfo {author} {\bibfnamefont {S.~V.}\ \bibnamefont {Dubonos}},
		\bibinfo {author} {\bibfnamefont {J.~G.~S.}\ \bibnamefont {Lok}}, \bibinfo
		{author} {\bibfnamefont {I.~V.}\ \bibnamefont {Grigorieva}}, \bibinfo
		{author} {\bibfnamefont {J.~C.}\ \bibnamefont {Maan}}, \bibinfo {author}
		{\bibfnamefont {L.~T.}\ \bibnamefont {Hansen}}, \ and\ \bibinfo {author}
		{\bibfnamefont {P.~E.}\ \bibnamefont {Lindelof}},\ }\bibfield  {title}
	{\enquote {\bibinfo {title} {Ballistic hall micromagnetometry},}\ }\href
	{\doibase 10.1063/1.120034} {\bibfield  {journal} {\bibinfo  {journal}
			{Applied Physics Letters}\ }\textbf {\bibinfo {volume} {71}},\ \bibinfo
		{pages} {2379} (\bibinfo {year} {1997})}\BibitemShut {NoStop}%
	\bibitem [{\citenamefont {Novoselov}\ \emph {et~al.}(2003)\citenamefont
		{Novoselov}, \citenamefont {Morozov}, \citenamefont {Dubonos}, \citenamefont
		{Missous}, \citenamefont {Volkov}, \citenamefont {Christian},\ and\
		\citenamefont {Geim}}]{Novoselov2003}%
	\BibitemOpen
	\bibfield  {author} {\bibinfo {author} {\bibfnamefont {K.~S.}\ \bibnamefont
			{Novoselov}}, \bibinfo {author} {\bibfnamefont {S.~V.}\ \bibnamefont
			{Morozov}}, \bibinfo {author} {\bibfnamefont {S.~V.}\ \bibnamefont
			{Dubonos}}, \bibinfo {author} {\bibfnamefont {M.}~\bibnamefont {Missous}},
		\bibinfo {author} {\bibfnamefont {A.~O.}\ \bibnamefont {Volkov}}, \bibinfo
		{author} {\bibfnamefont {D.~A.}\ \bibnamefont {Christian}}, \ and\ \bibinfo
		{author} {\bibfnamefont {A.~K.}\ \bibnamefont {Geim}},\ }\bibfield  {title}
	{\enquote {\bibinfo {title} {Submicron probes for hall magnetometry over the
				extended temperature range from helium to room temperature},}\ }\href
	{\doibase 10.1063/1.1576492} {\bibfield  {journal} {\bibinfo  {journal} {J.
				Appl. Phys.}\ }\textbf {\bibinfo {volume} {93}},\ \bibinfo {pages} {10053}
		(\bibinfo {year} {2003})}\BibitemShut {NoStop}%
	\bibitem [{\citenamefont {{Le Roy}}\ \emph {et~al.}(2016)\citenamefont {{Le
				Roy}}, \citenamefont {Shaw}, \citenamefont {Haettel}, \citenamefont
		{Hasselbach}, \citenamefont {Dumas-Bouchiat}, \citenamefont {Givord},\ and\
		\citenamefont {Dempsey}}]{Roy2016}%
	\BibitemOpen
	\bibfield  {author} {\bibinfo {author} {\bibfnamefont {D.}~\bibnamefont {{Le
					Roy}}}, \bibinfo {author} {\bibfnamefont {G.}~\bibnamefont {Shaw}}, \bibinfo
		{author} {\bibfnamefont {R.}~\bibnamefont {Haettel}}, \bibinfo {author}
		{\bibfnamefont {K.}~\bibnamefont {Hasselbach}}, \bibinfo {author}
		{\bibfnamefont {F.}~\bibnamefont {Dumas-Bouchiat}}, \bibinfo {author}
		{\bibfnamefont {D.}~\bibnamefont {Givord}}, \ and\ \bibinfo {author}
		{\bibfnamefont {N.~M.}\ \bibnamefont {Dempsey}},\ }\bibfield  {title}
	{\enquote {\bibinfo {title} {Fabrication and characterization of polymer
				membranes with integrated arrays of high performance micro-magnets},}\ }\href
	{\doibase 10.1016/j.mtcomm.2015.12.004} {\bibfield  {journal} {\bibinfo
			{journal} {Mater. Today Commun.}\ }\textbf {\bibinfo {volume} {6}},\ \bibinfo
		{pages} {50--55} (\bibinfo {year} {2016})}\BibitemShut {NoStop}%
	\bibitem [{\citenamefont {Dempsey}\ \emph {et~al.}(2014)\citenamefont
		{Dempsey}, \citenamefont {{Le Roy}}, \citenamefont {Marelli-Mathevon},
		\citenamefont {Shaw}, \citenamefont {Dias}, \citenamefont {Kramer},
		\citenamefont {Cuong}, \citenamefont {Kustov}, \citenamefont {Zanini},
		\citenamefont {Villard}, \citenamefont {Hasselbach}, \citenamefont {Tomba},\
		and\ \citenamefont {Dumas-Bouchiat}}]{Dempsey2014}%
	\BibitemOpen
	\bibfield  {author} {\bibinfo {author} {\bibfnamefont {N.~M.}\ \bibnamefont
			{Dempsey}}, \bibinfo {author} {\bibfnamefont {D.}~\bibnamefont {{Le Roy}}},
		\bibinfo {author} {\bibfnamefont {H.}~\bibnamefont {Marelli-Mathevon}},
		\bibinfo {author} {\bibfnamefont {G.}~\bibnamefont {Shaw}}, \bibinfo {author}
		{\bibfnamefont {A.}~\bibnamefont {Dias}}, \bibinfo {author} {\bibfnamefont
			{R.~B.~G.}\ \bibnamefont {Kramer}}, \bibinfo {author} {\bibfnamefont {L.~V.}\
			\bibnamefont {Cuong}}, \bibinfo {author} {\bibfnamefont {M.}~\bibnamefont
			{Kustov}}, \bibinfo {author} {\bibfnamefont {L.~F.}\ \bibnamefont {Zanini}},
		\bibinfo {author} {\bibfnamefont {C.}~\bibnamefont {Villard}}, \bibinfo
		{author} {\bibfnamefont {K.}~\bibnamefont {Hasselbach}}, \bibinfo {author}
		{\bibfnamefont {C.}~\bibnamefont {Tomba}}, \ and\ \bibinfo {author}
		{\bibfnamefont {F.}~\bibnamefont {Dumas-Bouchiat}},\ }\bibfield  {title}
	{\enquote {\bibinfo {title} {Micro-magnetic imprinting of high field gradient
				magnetic flux sources},}\ }\href {\doibase 10.1063/1.4886375} {\bibfield
		{journal} {\bibinfo  {journal} {Appl. Phys. Lett.}\ }\textbf {\bibinfo
			{volume} {104}},\ \bibinfo {pages} {262401} (\bibinfo {year}
		{2014})}\BibitemShut {NoStop}%
\end{thebibliography}

%

\end{document}